\documentclass[
	aps, prd, twocolumn, 
	10pt, notitlepage, a4paper,
        floats, floatfix,
	amsmath, amssymb, amsfonts, 
	superscriptaddress,
	showpacs, showkeys,
	nofootinbib
]{revtex4-1}
 \usepackage{amsmath,amssymb,amsfonts}
 \usepackage{graphics,subfigure}
 \usepackage[dvips]{graphicx,color}
 \usepackage[usenames]{xcolor}

\newcommand{\Tr}{\mbox{Tr}}

\newcommand{\R}{\mathcal {R}}

\def\Lie{\mathcal{L}}
\def\A{\mathcal{A}}
\def\X{\tilde{X}}
\def\p{\partial}

\begin{document}
 
\title{The Initial Value Formulation of Dynamical Chern-Simons Gravity}

\author{T\'erence Delsate}
\email{terence.delsate@umons.ac.be}
\affiliation{Theoretical and Mathematical Physics Department,
University of Mons, 7000 Mons, Belgium}

\author{David Hilditch}
\email{david.hilditch@uni-jena.de}
\affiliation{Theoretical Physics Institute, University of Jena, 07743 Jena,
Germany}

\author{Helvi Witek}
\email{h.witek@damtp.cam.ac.uk}
\affiliation{Department of Applied Mathematics and Theoretical Physics, 
Centre for Mathematical Sciences, University of Cambridge, Wilberforce 
Road, Cambridge CB3 0WA, UK}

\date{\today}

\begin{abstract}
We derive an initial value formulation for dynamical Chern-Simons gravity, a 
modification of general relativity involving parity-violating higher derivative 
terms. We investigate the structure of the resulting system of partial 
differential equations thinking about linearization around arbitrary backgrounds. 
This type of consideration is necessary if we are to establish well-posedness of 
the Cauchy problem. Treating the field equations as an effective field theory we 
find that weak necessary conditions for hyperbolicity are satisfied. For the full 
field equations we find that there are states from which subsequent evolution 
is not determined. Generically the evolution system closes, but is not hyperbolic in any 
sense that requires a first order pseudo-differential reduction.
In a cursory mode analysis we find that the equations of motion contain terms that may 
cause ill-posedness of the initial value problem.
\end{abstract}


\maketitle

\tableofcontents

\section{Introduction}

General relativity (GR) is the most successful theory of gravity to
date, and has  passed all experimental tests with flying colors.
However, these tests, such as observations of pulsar binaries or
observations inside the Solar System,  are restricted to the range in
which low order post-Newtonian calculations accurately describe the
dynamics~\cite{Will:2014kxa,Gair:2012nm,Sathyaprakash:2009xs,Antoniadis:2013pzd}.
Bearing in mind the extrapolation of GR over many orders of magnitude,
and the issues in wedding gravity with quantum physics, it would not
be surprising  if modifications to GR in the high curvature regime
were discovered.  Identifying {\it how} the field equations might be
modified is however open to debate.  One class of modifications  is
motivated by string theory in the low energy limit.  Specifically,
compactifications of~$10$-dimensional heterotic string theory to four
spacetime dimensions yield modifications of the Einstein-Hilbert
action  involving higher derivative terms of the
metric~\cite{polchinsky:string,Alexander:2009tp,Adak:2008yg}. The
gravity  sector of the action including quadratic terms in the
curvature
is~\cite{polchinsky:string,Alexander:2009tp,Yunes:2011we,Yunes:2013dva}
\begin{align}
\label{eq:GeneralAction}
S = & \int d^{4}x \sqrt{-g}\left[ \kappa R + f_1(\theta) R^2 +
  f_2(\theta) R_{ab} R^{ab}  \right.\nonumber\\ & \left.  +
  f_3(\theta) R_{abcd} R^{abcd} + f_4(\theta) \,^{\ast}\!R R
  \right.\nonumber\\ & \left.   -
  \frac{b_{\rm{CS}}}{2}\left(\nabla_{a}\theta\nabla^{a}\theta + 2V(\theta)
  \right) + \mathcal{L}_{\rm{m}}  \right]\,,
\end{align}
where the first term relates to GR with the gravitational coupling
constant~$\kappa$, $\mathcal{L}_{\rm{m}}$ denotes the Lagrangian for
ordinary matter, $\theta$ is a  dynamical scalar field, and
$f_i(\theta)$ are functions specifying the coupling  of the higher
derivative contributions.

With the specific choice~$f_1(\theta)=a_{\rm{GB}} \exp(-2\theta)$,
$f_2(\theta)=-4f_1(\theta)$,  $f_3(\theta)=f_1(\theta)$,
$f_4(\theta)=0$ we obtain the well known  dilaton-Gauss-Bonnet
modification with the coupling~$a_{\rm{GB}}$~\cite{Gross:1986mw}.  Its
parity violating counterpart includes the Pontryagin density
$\,^{\ast}\!R R =
\,^{\ast}\!R^{abcd}\,R_{bacd}=-\tfrac{1}{2}\epsilon^{cd}{}_{ef}
R^{abef} R_{abcd}$  with an axionic-type coupling to the scalar field
$\theta$, i.e., $f_1(\theta)=f_2(\theta)=f_3(\theta)=0$,
$f_4(\theta)=\tfrac{a_{\rm{CS}}}{4}\theta $ and is called dynamical
Chern-Simons (dCS)  theory~\cite{Jackiw:2003pm,Alexander:2009tp}.  If
the kinematic term is discarded, the resulting model is called
non-dynamical Chern-Simons theory.  If the scalar field  is constant
the corrections are topological in four dimensions and the equations
of motion reduce to those of GR. 

Some solutions of GR are inherited by the dCS model. Specifically, even
parity spacetimes, such as the Schwarzschild solution, have vanishing
Pontryagin density and are unaffected by the dCS modification.  In
contrast, the Kerr black hole (BH) is parity odd and therefore not a
solution in dCS gravity.  No complete solution for a rotating BH in
dCS theory is known, but see
Refs.~\cite{Yunes:2009hc,Yagi:2012ya,Yunes:2007ss,Molina:2010fb,Delsate:2011qp,Pani:2011gy,Konno:2014qua,Stein:2014xba}
for perturbative calculations. 
Exploring dynamical BH solutions provides the possibility to explore  
gravity in the strong-field  regime. In this
context modifications to GR may become important.  For example,
studies of extreme-mass-ratio inspirals in dCS gravity  revealed  an
additional polarization of gravitational waves (GWs)~\cite{Gair:2011ym,Sopuerta:2010ha,Sopuerta:2009iy,Canizares:2012is,Pani:2011xj}.
Furthermore rotating spacetimes are deformed in comparison to GR and may cause 
deviations in the GW signals.
These~``smoking-gun'' effects may be observable with
future  space-based GW detectors along the lines of
the eLISA mission~\cite{AmaroSeoane:2012km,Seoane:2013qna} or, in case
of solar-mass BH binaries,  with existing or upcoming ground-based
GW detectors such as the advanced  LIGO/ VIRGO
detector network~\cite{Abbott:2007kv,Acernese:2008zzf,
  Abadie:2011kd,Aasi:2013wya,aLIGO} or the KAGRA
detector~\cite{Aso:2013eba,Somiya:2011np}. 
GW astronomy might furthermore yield
more  stringent bounds on the dCS coupling
parameter~\cite{Yunes:2013dva}, which so far has been constrained by
table-top experiments~\cite{Yagi:2012ya} and observations of
frame-dragging effects in the Solar system~\cite{AliHaimoud:2011fw} to
be $\sqrt{|a_{\rm{CS}}|} \lesssim 10^8 km$.  Recently, it has been
suggested that observations of highly spinning, solar-mass BHs could
be employed to improve this bound to $\sqrt{|a_{\rm{CS}}|} \lesssim
\mathcal{O}(10)km$~\cite{Stein:2014xba}.

Investigating dCS gravity for comparable-mass binary systems, the
most promising sources for ground-based GW detectors, is still outstanding
in the high curvature regime (see Ref.~\cite{Yagi:2011xp} for a study in the PN approximation)
 -- missing is a formulation which could be treated by standard 
numerical relativity techniques~\cite{Alcubierre:2008}. It was foreseen in
Ref.~\cite{Garfinkle:2010zx} that the  higher derivative equations
might make such a formulation problematic.
Given the ease in prescribing modifications to GR
compatible with observational bounds it is natural to ask,
what other tests could we subject the modified theory to? An obvious
option is to look for logical  inconsistencies, or for contradictions
with some physical principle that we hold dear. In the  present work
we follow this tack. A fundamental question for any
field  theory is whether it has a locally well-posed initial value
problem (IVP). We might furthermore  insist on causality, or finite
propagation speeds of information.  
While the linear stability of specific solutions has been
studied~\cite{Cardoso:2009pk,Molina:2010fb,Garfinkle:2010zx,Ayzenberg:2013wua,Konno:2014qua}
(see Refs.~\cite{Izumi:2014loa} and~\cite{Reall:2014pwa} 
for similar studies in Gauss-Bonnet and in Lovelock theories),
it is not known whether or not
dCS gravity makes sense as a time evolution system.

Therefore, we perform a~$3+1$ decomposition of the dCS field equations along 
the lines of the  ADM--York split~\cite{Arnowitt:1962hi,York:1979,Gourgoulhon:2007ue}
and begin studying the structure of the resulting system of partial differential equations (PDEs).
Guided  by the similarities between the PDE structure of GR and
Maxwell's theory we first investigate  the properties of Maxwell's
equations modified by the Chern-Simons term coupled to an axion field,
which at first sight appears to be the electromagnetic analogue of dCS
gravity. 
In contrast to Maxwell-Chern-Simons theory,
we find that dCS gravity cannot be  written in first order form,
a necessary condition in many definitions of 
hyperbolicity~(see for example~\cite{NagOrtReu04}). Thus the attempted analysis 
fails: dCS gravity does not satisfy these definitions, even so far as GR does 
before fixing the gauge.
Thus the naive expectation would be that even if the 
field equations admit a well-posed initial value problem, signals may travel 
arbitrarily fast. In any relativistic theory, however, physical signals, to 
be contrasted with gauge, should propagate at finite speeds.

dCS gravity is not normally viewed as a complete
theory, but rather as  an {\it effective} theory, emerging as a higher
derivative modification to GR in string theory, loop
quantum gravity~\cite{Ashtekar:1988sw,Mercuri:2009zt,Taveras:2008yf},
cosmological inflation~\cite{GarciaBellido:2003wd} or particle
physics~\cite{Mariz:2004cv}. 
The effective theory is a reasonable model when its solutions are a good 
approximation to those of the full theory. An approach in the literature is 
either to reduce the order of the highest derivative assuming a small coupling  
or to treat the effective field equations order by order in the coupling 
parameter. The resulting PDEs can be reduced to first order, thus fulfilling 
this very weak requirement to have a chance to be hyperbolic.

The paper is structured as follows. In
Sec.~\ref{section:Hyperbolicity},  we review relevant aspects of PDE theory. 
In Sec.~\ref{section:dCS_Maxwell} we discuss the Chern-Simons modification  
to the Maxwell equations. Subsequently, in Sec.~\ref{section:dCS} we  present 
the full dCS field equations in~$3+1$ decomposed form.
In Sec.~\ref{section:Effective} we  discuss
how some of the problems we encounter can be avoided when the model
is treated as an effective theory. Finally, in Sec.~\ref{section:Conclusions}, 
we conclude. We use geometrized
units~$G=1=c$ throughout. Early lower letters $a,b,\ldots
\in{0,\ldots,3}$  denote spacetime indices; middle lower letters
$i,j,\ldots\in{1,2,3}$  denote spatial indices.

\section{Hyperbolic and Parabolic PDEs}\label{section:Hyperbolicity}
Because the structure and properties of time evolution PDEs play a central role in the
present paper we begin with a brief discussion highlighting the difference between hyperbolic
and parabolic PDEs.

Any reasonable physical model should result in PDE problems that  are
well-posed. Roughly speaking, well-posedness is the existence  of a
unique solution which depends continuously on given initial data. 
In a relativistic context we additionally insist on finite propagation
speeds for physical fields, as opposed to gauge, given arbitrary data. 
Hyperbolic PDEs  are characterized by this property. Formal definitions 
of hyperbolicity are given for first order systems in terms of algebraic 
properties of the coefficients of the derivatives~\cite{KreLor89,Gustafsson1995}.
Hyperbolicity of higher order derivative systems is defined by
considering  properties of fully first order (pseudo-) differential
reductions~\cite{Tay81,NagOrtReu04,GunGar05,Hilditch:2014naa}. 
Therefore a necessary condition for the application of these definitions 
is the existence of a first order reduction of the PDE system in 
question.
In the remainder of the paper we will refer to this definition 
of hyperbolicity without further comment.
Consider the linear, constant coefficient, first order  in time, second
order in space (FT$2$S) system, 
\begin{align}
\p_t u&=(A^u{}_u)^i\p_i u+(A^u{}_v) v+S_{u},\nonumber\\ \p_t
v&=(A^v{}_u)^{ij}\p_i\p_j u +(A^v{}_v)^i\p_i
v+S_{v}.\label{eq:systems_FOITSOIS}
\end{align}
This system can be reduced to first differential order by introducing
the variables~$d_i=\p_iu$, and then appropriately adding the
constraint~$c_i=d_i-\p_iu$ to the resulting equations.
In general, we will call equations with this shape~``FT$N$S'', which stands for
{\it first order
 in time, $N$-th order in space.}  Specifically, the dCS equations of
motion (EoMs) contain third derivatives of the metric, so we might like to
end up with an~FT$3$S PDE system, 
\begin{align}
\nonumber \p_t u &= (A^u{}_{u})^i \p_i u +(A^u{}_{v}) v
+S^u,\\ \nonumber \p_t v &= (A^v{}_{u})^{ij} \p_i\p_j u +(A^v{}_{v})^i
\p_i v +(A^v{}_{w}) w+S^v,\\ \nonumber \p_t w &= (A^w{}_{u})^{ijk}
\p_i\p_j\p_k u +(A^w{}_{v})^{ij} \p_i\p_j v\\ &\quad+(A^w{}_{w})^{i}
\p_i w + S^w,\label{eq:FT3S}
\end{align}
which is easily seen~\cite{Hilditch:2014naa} to be the natural
generalization  of Eq.~\eqref{eq:systems_FOITSOIS}.

The archetypal hyperbolic PDE is the wave equation, which can be
written with a first order in time reduction as
\begin{align}
\p_t\Phi(t,x)&=\Pi(t,x)\,,\qquad \p_t\Pi(t,x)=\p^2_x\,\Phi(t,x)\,.
\end{align}
The fundamental solution of the wave equation, that is the response to
a Dirac delta function placed at the origin initially, is
\begin{align}
\Phi(t,x) &= \frac{1}{4}\,\Theta(t) \big[ \mbox{Sign}(t+x) +
  \mbox{Sign}(t-x) \big],
\end{align}
where~$\Theta$ is the Heaviside function.  For every $t>0$ the
solution to the wave equation~$\Phi$ has a compact support in~$x$.
Plotting the fundamental solution to the wave equation shows that the
evolving pulse remains at all times inside the future null cone of the
initial pulse.  Contrast this with the heat equation,
\begin{align}
\p_t\Phi(t,x)&=\p^2_x\,\Phi(t,x)\,.
\end{align}
Introducing here a reduction variable~$d_x=\p_x\Phi$ does not reduce
the PDE to first  order because the equation generates terms
like~$\p^2_xd_x$. The fundamental solution  in this case is given by,
\begin{align}
\Phi(t,x) = \frac{1}{2\sqrt{\pi t}}\Theta(t)e^{-\frac{x^2}{4t}}\,.
\end{align}
For every $t>0$ the solution to the heat equation has an infinite
support. This means  that a point impulse propagates instantaneously
everywhere once~$t>0$~\cite{evansPDE}. Similar statements can be made
about Schr\"odinger like equations. This `causality violation'
property is present in other parabolic PDEs and is not permissible  in
relativistic physics where we have a natural speed limit. 

The discussion so far is relevant for linear PDEs. When facing
non-linear problems we  must linearize the equations about an
arbitrary solution, and apply the linear theory.  For certain types 
of equations, such as hyperbolic or parabolic, and if certain
smoothness conditions are satisfied~\cite{KreLor89} then
well-posedness of the  linear problem guarantees {\it local in time}
well-posedness of the non-linear problem.  It is possible that the
local classification changes over the domain, the tricomi
equation~$\p_x^2u=x\p_y^2u$ being the standard example of this
behavior. We will also see that it is possible that, in some region, 
the PDE does not fall into any of the standard classes. In this case 
a more ad hoc analysis may be all that is available.

\section{Chern-Simons electromagnetism}\label{section:dCS_Maxwell}

\subsection{Action and field equations}

Bearing in mind the similarities between the PDE structure of GR and
electromagnetism it is instructive to investigate the hyperbolicity
properties of Maxwell's equations modified by a Chern-Simons term,
which we will call Chern-Simons electromagnetism, before turning to 
the gravity case. 

The action consists of an axionic deformation of the standard
electromagnetic action~\cite{Arvanitaki:2009fg}. The corresponding
Lagrangian density is given by
\begin{align}
\label{eq:MCSaction}
\mathcal L_{\rm{CSE}}&= -\frac{1}{4}F^{ab} F_{ab} -
\frac{\lambda}{2}\psi\, ^{\ast}\!F_{ab} F^{ab}
-\frac{1}{2}\nabla^{a}\psi\nabla_{a}\psi - V(\psi)\,,
\end{align}
where~$F_{ab}=\nabla_{a}A_{b}-\nabla_{b}A_{a}$ is the field strength,
$^\ast\!F_{ab} = \tfrac{1}{2}\epsilon_{ab}{}^{cd}F_{cd}$ is its dual,
$A^a$ is the $U(1)$ gauge field  and $\lambda$ denotes the coupling to
the scalar field $\psi$.  The term $\,^{\ast}\!F_{ab} F^{ab}$ imposes
the parity violation and can be interpreted as the analogue to the
Pontryagin density in the  gravity case. The resulting EoMs are
\begin{align}
\nabla^{a}\nabla_{a}\psi &= \tfrac{1}{2}\lambda\,^{\ast}\!F_{ab}
F^{ab} + V'(\psi) \,,\nonumber\\ \nabla_{b} F^{ab} &= -2 \lambda
\,^{\ast}\!F^{ab}\nabla_{b}\psi \,,\quad \nabla_{b}\,^{\ast}\!F^{ab} =
0\,.\label{eq:MCSeom4d}
\end{align}
Note, that the last relation is satisfied trivially when expressing it 
in terms of the vector potential and we only keep it for completeness.
Already at this level, we observe that this system of PDEs can be made
strongly hyperbolic because in the appropriate gauge it consists of a
set  of decoupled wave equations for the scalar and gauge fields,
respectively, with some lower order source terms. In a PDE analysis
language, this system is said to be minimally coupled.

\subsection{3+1 decomposition}

In this section, we show explicitly that Eqs.~\eqref{eq:MCSeom4d}  are
indeed minimally coupled and rewrite them as a FT$2$S system.
Therefore, we foliate a $4$-dimensional spacetime
manifold~$\mathcal{M}$  into $3$-dimensional spatial
hypersurfaces~$\Sigma_{t}$ parametrized by  the coordinate
time~$t$. We denote the spatial metric~$\gamma_{ij}$ and the unit
timelike vector~$n^a$ which is orthogonal to the spatial slices and
satisfies~$n_a n^a = -1$. Furthermore, we introduce the  projection
operator $\gamma_a{}^b = \delta_a{}^b + n_a n^b$. Within  this
decomposition the spacetime line element is 
\begin{align}
\label{eq:lineelement3+1}
ds^2 = & -\alpha^2 dt^2 + \gamma_{ij} (dx^i + \beta^i dt)  (dx^j +
\beta^j dt)\,,
\end{align}
where~$\alpha$ and~$\beta^i$ are the lapse function and shift vector.

To rewrite Eqs.~\eqref{eq:MCSeom4d} as a time evolution problem, we
decompose  the~$4$-vector potential~$A_{a}$ into its spatial
part~$\A_{i}$ and normal  component~$\Phi$ with a convenient
normalization, according to,
\begin{align}
\label{eq:MCSdefAphi}
A_a  = & \A_a + \frac{\Phi}{\alpha} n_a \,,\quad  \A_i =
\gamma^{a}{}_{i} A_{a} \,,\quad  \Phi = - \alpha\, n^a A_a\,.
\end{align}
Next, we introduce the electric and magnetic fields~$E_{a}$ and
$B_{a}$  given by the contractions of the Maxwell tensor and its dual
with the unit  timelike vector, 
\begin{align}
\label{eq:MCSdefE}
E_{i} = & \gamma^{a}{}_{i} F_{ab} n^{b}\,,\quad  B_{i} =
\gamma^{a}{}_{i} \,^\ast\!F_{ab} n^b\,.
\end{align}
The magnetic field is related to the spatial component of the  vector
potential via
\begin{align}
\label{eq:BinAEM}
B_{i} = & \epsilon_{i}{}^{jk}D_{j}\A_{k}\,.
\end{align}
The magnetic field $B_{i}$ is not treated as a dynamical
variable itself. We employ it {\it purely as a shorthand}
whenever economical.

Given these relations we can re-express the Maxwell tensor in terms of
the  electric field and the spatial vector potential,
\begin{align}
F_{ab}          &=  n_a E_b - n_bE_a + D_a\A_b - D_b\A_a
\,,\nonumber\\ \,^{\ast}\!F_{ab} &= n_{a} B_{b} - n_{b} B_{a} +
\epsilon_{abc} E^{c} \,,\label{eq:MCSFinAE}
\end{align}
where~$D_{i}$ denotes the~$3$-dimensional covariant derivative
associated with the spatial metric $\gamma_{ij}$ and the analogue of
the Pontryagin density is~$\,^{\ast}\!F^{ab}F_{ab} = -4 E_{i} B^{i}$.
We introduce the reduction variable~$\Pi_{\psi}$, 
\begin{align}
\Pi_{\psi} := - n^a\nabla_a\psi = -
\frac{1}{\alpha}(\p_{t}-\Lie_{\beta})\psi\,.
\end{align}
Then, employing the~$3+1$ decomposition, we obtain a set of time
evolution equations,
\begin{align}
\p_t \psi &= - \alpha \Pi_{\psi} +
\Lie_{\beta}\psi\,,\nonumber\\ \p_t\A_i  &=  - \alpha E_i - D_i\Phi +
\Lie_{\beta}\A_i \,,\nonumber\\ \p_t \Pi_{\psi} &=  \alpha[ K
  \Pi_{\psi}+V'(\psi)  - D^i D_i\psi + 2\lambda E_i B^i]
\nonumber\\ &\quad  - D_{i}\psi D^{i}\alpha+ \Lie_{\beta} \Pi_{\psi}
\,,\nonumber\\ \p_t E^i &=   D^{j}\Big[\alpha(D^i\A_j-D_j\A^i)\Big] +
\alpha K E^i  - 2 \lambda \alpha \Pi_{\psi} B^i \nonumber\\ &\quad  -
2 \lambda \alpha\,[E\times (D\psi)]^i  + \Lie_{\beta}
E^i\,,\label{eq:MCSEvol}
\end{align}
with the cross product defined by,
\begin{align}
[E\times(D\psi)]^i = & \epsilon^{ijk} E_{j} D_{k}\psi\,.
\end{align}
For completeness we give also the time derivative of the magnetic
field,
\begin{align}
\p_tB^i&=\alpha K B^i-[D\times (\alpha E)]^i+\Lie_\beta B^i\,.
\end{align}
This equation is independent of any particular choice of model for the
electromagnetic field, because it follows directly
from~$\nabla_{b}\,^{\ast}\!F^{ab}=0$. Finally, the constraint equation
reads
\begin{align}
\label{eq:MCSCon}
M &=  D^{i}E_{i} - 2 \lambda B^{i} D_{i}\psi = 0\,.
\end{align}
We may formally compute the time derivative of the constraint, and
find,
\begin{align}
\p_tM&= \alpha K M+\Lie_\beta M\,.\label{eq:EM_M}
\end{align}
As expected the constraint subsystem is closed, i.e. if the constraint
is satisfied initially, it is satisfied during the entire evolution.
If we were going to analyze hyperbolicity of a particular formulation
of the theory we would now adopt the free-evolution point of view,
make  a choice of gauge for the field~$\Phi$, expand the solution
space with  new constraints and couple them to the present
system~\cite{Hilditch:2013sba}. Instead we will focus simply on the
structure of the equations. 

We may view the model equations as telling us one
constraint~\eqref{eq:EM_M} and the three evolution equations
for~$E^i$. The  remaining equations are differential identities
following from the fact that~$F_{ab}$ is a closed two-form. It is
useful to keep this in mind  in the gravitational case that follows.

Looking again at the Maxwell-Chern-Simons field equations expressed
in a first order in time form, we see that the sets~$(\A_i,E^i)$
and~$(\psi,\Pi_\psi)$ are minimally coupled. More precisely, the
system~\eqref{eq:MCSEvol} has the~FT$2$S structure given in 
Eq.~\eqref{eq:systems_FOITSOIS}, where $\A_{i},\ \psi$ are $u$-like
variables and $E^i,\ \Pi_\psi$ are $v$-like. In an appropriate
formulation~$\Phi$ will be a $u$-like variable. Furthermore the block
of the principal symbol  associated with the first pair
$(\A_{i},E^{i})$  is identical to that of the pure Maxwell equations,
and that of the second pair $(\psi,\Pi_{\psi})$ to that of the wave
equation. In  other words, after a suitable gauge choice, the full
system can be  rendered strongly hyperbolic according to a treatment
identical to that  for the Maxwell equations~\cite{Hilditch:2013ila}.

\section{Chern-Simons gravity}
\label{section:dCS}

\subsection{Motivation from String Theory}\label{sec:mst}

Extensions of GR involving dCS modifications are motivated for example
by the compactification of the bosonic part of $10$-dimensional
heterotic string theory to $4$-dimensional  $\mathcal{N}=1$
Supergravity~\cite{polchinsky:string}.  The bosonic sector of this
theory is given by
\begin{align}
S_{10D} &= \int d^{10} x \sqrt{-g_{10}} \Bigl[ R -
  \frac{1}{2}\partial_a\phi\,\partial^a\phi - \frac{1}{12} e^{-\phi}
  H_{abc} H^{abc} \nonumber\\ &- \frac{1}{4}e^{-\frac{\phi}{2}} \Tr
  \left( F_{ab} F^{ab} \right) \Bigr],
\end{align}
where $g_{10}$ is the $10$-dimensional metric, $F$ and $H$ are $2$-
and $3$-form field strengths, respectively, and $\phi$ is a scalar
field. 

It was shown that GR in even dimensions suffers from a gravitational
anomaly~\cite{alvarez:1984}, which can be cured by shifting the
$3$-form field strength with additional Chern-Simons terms, as shown
by Green and Schwarz \cite{schwartz:1987, Schwartz:1987b}
\begin{align}
H_3 = & dB_2 -\tfrac{1}{4} \left(\Omega_3(A) - \Omega_3(\omega)
\right) \,,
\end{align}
where $B_2$ is a $2$-form, $A$ is the Yang-Mills $1$-form and $\omega$
is the spin connection.

The terms involving $\Omega_3$ are obtained from the Green-Schwarz
prescription and are defined as
\begin{align}
\label{eq:ChernForm}
\Omega_3(A) = & Tr\left( dA\wedge A +\tfrac{2}{3}A \wedge A\wedge A
\right).
\end{align}
Here, it is assumed that all moduli except for the axion are
stabilized, and the resulting $4$-dimensional action is that of dCS
theory. See the review \cite{Alexander:2009tp} for details and
references.

For the $1$-form~$A$ the Chern-Simons form~\eqref{eq:ChernForm}
produces at most terms of the order~$(\p A)^2$ yielding an action that
does not involve higher derivative terms. In
Sec.~\ref{section:dCS_Maxwell} we  saw that for electromagnetism,
which has the same PDE structure as Yang-Mills,  Chern-Simons like
terms coming from the anomaly canceling procedures are structurally
fine. Thus one expects that, with a little work, dCS theory for
a~$1$-form admits a  well-posed initial value formulation. This
picture changes dramatically if we consider  the gravitational
sector. Bearing in mind that the spin connection behaves
as~$\omega\sim\p g_{10}$, it is evident that the Chern
form~\eqref{eq:ChernForm} will  introduce a term of the form~$\p^2
g_{10} $ leading to an action which contains  higher derivatives of the
metric.  We will show in the remainder of this section that these
higher derivative terms  prevent dCS gravity from being hyperbolic.

\subsection{Action and field equations}

We now focus our attention to the case of Chern-Simons gravity coupled
to a dynamical scalar field. We recover the corresponding action by
setting~$f_1(\theta)=f_2(\theta)=f_3(\theta)=0$
and~$f_4(\theta)=\tfrac{a_{\rm{CS}}}{4}\theta$ in
Eq.~\eqref{eq:GeneralAction} and note it here only for
completeness~\cite{polchinsky:string,Alexander:2009tp}
\begin{align}
\label{eq:ActiondCS}
S = & \int d^{4}x \sqrt{-g}\left[ \kappa R +
  \frac{a_{\rm{CS}}}{4}\theta\,\,^{\ast}\!R R  \right.\nonumber\\ &
  \left.  - \frac{b_{\rm{CS}}}{2}\left( \nabla^{a}\theta
  \nabla_{a}\theta + 2 V(\theta) \right) + \mathcal{L}_{\rm{m}}\right]
\,,
\end{align}
where~$\kappa$ is the gravitational coupling,~$a_{\rm{CS}}$ is the
axionic coupling of the scalar field~$\theta$ to the Pontryagin
density~$\,^{\ast}\!R R = -\tfrac{1}{2}\epsilon^{cd}{}_{ef} R^{abef}
R_{abcd}$ and~$ b_{\rm{CS}}$ denotes the coupling to the kinetic term
of the scalar  field. One recovers GR minimally coupled to a scalar
field if~$a_{\rm{CS}}=0$  and~$b_{\rm{CS}}=1$. From now on we will
consider the absence of ordinary  matter,
i.e.~$\mathcal{L}_{\rm{m}}=0$, and vanishing scalar field
potential~$V(\theta)=0$. If~$V(\theta)$ contains no derivatives of
the scalar field these assumptions will not change the outcome of the
hyperbolicity analysis. The EoMs are,
\begin{align}
G_{ab} + \frac{a_{\rm{CS}}}{\kappa} C_{ab} &=
\frac{b_{\rm{CS}}}{2\kappa}T^{\theta}_{ab} \,,\nonumber\\ \Box\theta
+ \frac{a_{\rm{CS}}}{4\,b_{\rm{CS}}} \,^{\ast}\!R R &= 0
\,,\label{eq:dCSEoM4D}
\end{align}
where~$G_{ab}$ is the Einstein tensor,~$T^{\theta}_{ab}$ is the
energy-momentum  tensor related to the scalar field,
\begin{align}
\label{eq:TTtheta}
T^{\theta}_{ab} = & \nabla_{a}\theta\nabla_{b}\theta  - \frac{1}{2}
g_{ab} \nabla^{c}\theta \nabla_{c}\theta\,,
\end{align}
and
\begin{align}
\label{eq:CtensorDCS}
C_{ab} = & \nabla_{c}\theta \epsilon^{cd}{}_{e(a}\nabla^{e}R^{}_{b)d}
+ \nabla^{c}\nabla^{d}\theta\,^{\ast}\!R_{d(ab)c}\,,
\end{align}
is the C-tensor. Already at this stage it is evident that there is a
distinction between the electromagnetic and gravitational Chern-Simons
models, since the scalar is not minimally coupled to the parent field
in the gravitational case.

In Refs.~\cite{Grumiller:2007rv,Garfinkle:2010zx} it was observed that
it is convenient  to rewrite the C-tensor in terms of the Weyl
tensor~$W_{abcd}$, resulting in
\begin{align}
\label{eq:CtensorinWeyl}
C_{ab} & = 2 \left(\nabla^{c}\theta\right)
\nabla^{d}\,^{\ast}\!W_{d(ab)c} +
\left(\nabla^{c}\nabla^{d}\theta\right) \,^{\ast}\!W_{d(ab)c}
\end{align}
where we have used the relation
\begin{align}
\label{eq:Bianchi_Weyl}
\nabla_{a}W^{abcd} = & \nabla^{[c}R^{d]b} +
\frac{1}{6}g^{b[c}\nabla^{d]}R\,,
\end{align}
that follows from the Bianchi identities (see, e.g.,
Ref.~\cite{Alcubierre:2008} and references therein). The strength of
this approach  is the fact that contractions of the Weyl tensor with a
timelike unit vector $n^{a}$  define its electric and magnetic parts
\begin{align}
\label{eq:EMdecompWeyl}
E_{ij} = & \gamma^{a}{}_{i} \gamma^{b}{}_{j} W_{acbd} n^{c}n^{d}
\,,\quad B_{ij} = \gamma^{a}{}_{i} \gamma^{b}{}_{j}
\,^{\ast}\!W_{acbd} n^{c}n^{d} \,.
\end{align}
in analogy with electromagnetism. Then, the Weyl tensor can be
reconstructed from its electric and magnetic
components~\cite{Alcubierre:2008}
\begin{align}
\label{eq:WeylinEB}
W_{abcd} = & 2\left(l_{a[c}E_{d]b} - l_{b[c}
  E_{d]a}-\epsilon^{e}{}_{ab} n_{[c}B_{d]e}  - \epsilon^{e}{}_{cd}
n_{[a}B_{b]e} \right)\,,
\end{align}
where~$l_{ab} = g_{ab} + 2 n_{a} n_{b}$. The Pontryagin density can be
written  as~$\,^{\ast}\!R R = -16 E_{ij} B^{ij}$~\cite{Grumiller:2007rv}.

\subsection{Formulation as Cauchy problem}

We proceed in our analysis by rewriting dCS gravity as a Cauchy
problem. For this we~$3+1$ decompose the EoMs~\eqref{eq:dCSEoM4D}.  
Although this is conceptually
straightforward it becomes rather  involved due to the presence of
covariant derivatives of the  Ricci tensor (yielding third derivatives
of the metric) in the  C-tensor; cf. Eq.~\eqref{eq:CtensorDCS}. The
computation was carried  out using the~\textsc{xTensor}~\cite{xtensor}
package.  For clarity we suppress some details of the derivation and
refer the interested reader to the notebooks~\cite{DMH_web}.
Some of  the relations we derive are purely of geometrical
origin and are,  therefore, {\textit{independent}} of the
gravitational field equations.  A second set of equations stem from
the EoMs and so are model dependent.
Specifically, the constraint equations all originate from various
projections of the  EoMs. Instead, the time evolution
equations  consist of both kinematical, i.e. geometric or model
independent, as  well as dynamical, i.e. model dependent, degrees of
freedom. A similar  decomposition was made
elsewhere~\cite{Shaun:2009}, but given in a less  geometric language
without employing~$E_{ij}$ and~$B_{ij}$, which unfortunately  gives
the impression that the constraint equations depend on the coordinate
gauge.

\vspace{10pt}
\paragraph*{Structure equations and choice of variables:}
The foliation of spacetime into $3$-dimensional spatial slices
introduces the spatial metric~$\gamma_{ij}$ together with the
extrinsic  curvature, 
\begin{align}
\label{eq:ExtrCurv}
K_{ij}= & -\frac{1}{2\alpha}(\p_t-\Lie_{\beta})\gamma_{ij}\,.
\end{align}
The spacetime coordinates are described by the lapse function $\alpha$
and shift  vector~$\beta^{i}$. The line element in terms of $3+1$
variables is  given in Eq.~\eqref{eq:lineelement3+1}.  
We introduce the reduction variable to the scalar field $\theta$,
\begin{align}
\label{eq:MCSdefKtheta}
\Pi & = - n^{a}\nabla_{a} \theta  = -
\frac{1}{\alpha}(\p_t-\Lie_{\beta})\theta \,.
\end{align}
It proves useful to split rank-2 tensors into their trace and
tracefree  parts. Specifically, we split the extrinsic
curvature~$K_{ij}$ and spatial Ricci  tensor~$\R_{ij}$ according to
\begin{align}
K_{ij} = & A_{ij} + \frac{1}{3}\gamma_{ij} K\,,\quad \R_{ij} =
\R^{\rm{TF}}_{ij} + \frac{1}{3}\gamma_{ij} \R\,,
\end{align}
The electric and magnetic parts of the Weyl tensor also enter the
equations  of motion. In~$3+1$ language their definitions give
\begin{align}
E_{ij} &=
\frac{1}{2}\R_{ij}^{\rm{TF}}+\frac{1}{2\alpha}[D_iD_j\alpha{}]^{\rm{TF}}
+\frac{1}{2}\Lie_nA_{ij} \nonumber\\ &\quad
+\frac{1}{6}KA_{ij}+\frac{1}{3}A^{kl}A_{kl}\gamma_{ij}\,,\nonumber\\ B_{ij}
&= (D\times A)_{ij}\equiv\epsilon_{(i|}{}^{kl}D_{k} A_{l|j)}\,.
\label{eq:Mag_Weyl}
\end{align}
Both quantities are already tracefree. While $E_{ij}$ joins the state
vector of dCS gravity as a dynamical variable the magnetic part will
be employed purely as a shorthand. The magnetic part satisfies the
geometric identities
\begin{align}
\label{eq:Mag_Weyl_div}
\p_tB_{ij} &= \alpha\Big[\,\big[D\times (2E - E^{\rm{GR}})\big]_{ij}
  -3A^k{}_{(i}B_{j)k}\nonumber\\ &\quad\quad
  -2\epsilon_{(i}{}^{kl}E_{j)k}D_l\ln\alpha
  -\epsilon_i{}^{kl}B_{km}A_{ln}\epsilon_j{}^{mn}
  \nonumber\\ &\quad\quad
  +\frac{1}{3}KB_{ij}+\frac{1}{2}\epsilon_{(i}{}^{kl}A_{j)k}
  M^{\rm{GR}}_l \Big] +\Lie_\beta B_{ij} \,, \nonumber \\ D^{j}B_{ij}
& = \epsilon_{i}{}^{jk} A^{l}{}_{k} E^{\rm{GR}}_{jl} + \frac{1}{2}
\epsilon_{i}{}^{jk} D_{j} M^{\rm{GR}}_{k} \,,
\end{align}
which follow from the projections of the Bianchi identities. 
Here, we use the shorthands,
\begin{align}
\label{eq:EGR_vac}
E_{ij}^{\rm{GR}}&=\R_{ij}^{\rm{TF}} - A^{k}{}_{i} A_{jk} +
\frac{1}{3}\gamma_{ij} A^{kl} A_{kl}  + \frac{1}{3}K A_{ij}\,,
\end{align}
for the expression that follows for the electric part of the  Weyl
tensor in {\it vacuum} GR, and
\begin{align}
\label{eq:GRmomentumconstraint}
M^{\rm{GR}}_i & = D^{j}A_{ij} - \frac{2}{3} D_{i}K
\,,
\end{align}
for the expression that appears in the vacuum momentum constraint  in
GR. In what follows we will also use the expression for the  vacuum
Hamiltonian constraint in GR,
\begin{align}
\label{eq:GRHamiltonian}
H^{\rm{GR}} & =\R - A_{ij}A^{ij} + \tfrac{2}{3}K^2 \,.
\end{align}
To summarize, at this stage the independent, dynamical variables are
taken to be~$(\gamma_{ij}, \theta, A_{ij}, K, E_{ij}, \Pi)$, while the
remaining quantities are used as shorthand notation.

\vspace{10pt}
\paragraph*{Auxiliary variables:}
For the sake of simplifying the expressions, we define
\begin{align}
\label{eq:DefOp}
X_{ij} = E_{ij} - E^{\rm{GR}}_{ij},\qquad \mathcal O_{ij}{}^{kl} =
\gamma_{(i}{}^{(k}\epsilon_{j)}{}^{l)m} D_m \theta,
\end{align}
and the auxiliary tensor
\begin{align}
\tilde{X}_{ij} = & \mathcal O_{ij}{}^{kl} X_{kl},
\end{align}
which will turn out to be an important object. We use the same
notation to denote the operator~$\mathcal{O}$ acting on other symmetric tensors.
This operator is not invertible, which can be checked by
explicitly computing its determinant in a particular basis.
It is also not nilpotent, see App.~\ref{app:On}, which is an important property for our purposes. 
In the model-dependent EoMs we will find a coupling to the gradient of
the scalar field $\theta$.  As long as this gradient is non-vanishing
it is useful to introduce a unit normal vector $s^{i}$ parallel to
$D^{i}\theta$ and, in particular, we define
\begin{align}
s_i = & L^{-1} D_i\theta \,,\qquad L^2 =  D_i\theta D^i \theta \,.
\end{align}
Furthermore, we introduce the $2$-metric $q_{ij}$ of the hypersurface
orthogonal to $s^{i}$, which defines a projection operator, and the
corresponding antisymmetric tensor
\begin{align}
q_{ij} &= \gamma_{ij} - s_i s_j \,,\qquad \epsilon_{jk} =
\epsilon_{ijk} s^i \,.
\end{align}
Then, the operator~$\mathcal{O}$ can be expressed as,
\begin{align}
L^{-1}\mathcal O_{ij}{}^{kl} = q_{(i}{}^{(l} \epsilon_{j)}{}^{k)} +
s_{(i} s^{(l} \epsilon_{j)}{}^{k)},
\end{align}
Using the projector,~$X_{ij}$ can be written as
\begin{align}
X_{ij}&=X_{ss} \big[s_i s_j-\tfrac{1}{2}q_{ij}\big] +
2X_{sA}q^A{}_{(i}
s_{j)}\nonumber\\ &\quad+X^{\textrm{TF}}_{AB}\big[q^A{}_{(i}q^B{}_{j)}-\tfrac{1}{2}q_{ij}q^{AB}
  \big ] ,\label{eq:XfromXtilde}
\end{align}
with,
\begin{align}
X^{\textrm{TF}}_{AB} &=
\big(q^i{}_{(A}q^j{}_{B)}-\tfrac{1}{2}q_{AB}q^{ij}\big)X_{ij}
\,,\nonumber\\  X_{sA} & = q_A{}^k s^l X_{kl} \,,\quad  X_{ss}  =
s^is^j X_{ij} \,,
\end{align}
where we use upper case latin indices to denote components that have
been  projected onto the $2$-surface and where indices $s$ refer to quantities contracted
with the normal vector $s^{i}$.  Then, 
\begin{align}
L^{-1}\tilde X_{ij} = & X^{\textrm{TF}}_{AC}\epsilon_B{}^{C}
q^A{}_{(i} q^B{}_{j)} + X_{sA} \epsilon_B{}^A q^B{}_{(i} s_{j)}.
\end{align}
The consequence of the operator~$\mathcal O$ being non-invertible is
that it is not possible to solve for all of the components of~$X_{ij}$
given~$\tilde{X}_{ij}$. Indeed, we can \emph{only} solve for~$X_{sA}$
and the projected tracefree part~$X^{\textrm{TF}}_{AB}$:
\begin{align}
X^{\textrm{TF}}_{AB} &= L^{-1} \tilde{X}_{ij} q^{i}{}_{C}
q{}^j{}_{(A}\epsilon^C{}_{B)} \,,\nonumber\\ X_{sA} &= 2 L^{-1}
\tilde{X}_{si} q^i{}_B\,\epsilon^B{}_{A}\,.
\end{align}
In conclusion, four of the five components of~$X_{ij}$ can be
expressed in terms of~$\tilde X_{ij}$. This is in big part -- but not
exclusively -- the main problem that arises in dCS gravity, as we will see in
the following. 
For convenience we will express all equations in terms of
$3$-dimensional, spatial variables, resorting to 
the $2+1$-split only where it is necessary for clarity.

\vspace{10pt}
\paragraph*{Constraint equations:}
The~$3+1$ split of the EoMs~\eqref{eq:dCSEoM4D} along
the lines of the ADM--York
decomposition~\cite{Arnowitt:1962hi,York:1979,Alcubierre:2008} yields
a set of constraint and time evolution equations. As in GR we obtain
scalar~$\mathcal{H}$ and vector constraints~$\mathcal{M}_{i}$ considering various projections of the EoMs.
Contracting the tensorial Eq.~\eqref{eq:dCSEoM4D} twice with the normal vector~$n^{a}$ yields
the  scalar constraint,
\begin{align}
\label{eq:dCSHamiltonian}
\mathcal{H} & = H^{\rm{GR}} - \frac{b_{\rm{CS}}}{2\kappa} \left( \Pi^2
+ D^{i}\theta D_{i}\theta \right) - \frac{2 a_{\rm{CS}}}{\kappa} \Big[
  2A^{ij}\tilde X_{ij} \nonumber \\ & \quad -B^{ij}\bigl( \Pi A_{ij}-
  D_{i}D_{j}\theta \bigr)  + (D\times M^{\rm{GR}})_{i}
  D^{i}\theta\Big]\,.
\end{align}
We obtain the vector constraint of the dCS gravity model by
considering the mixed projection of Eq.~\eqref{eq:dCSEoM4D}. The
computation gives
\begin{align}
\label{eq:dCSMomentum}
\mathcal{M}_{i} & = M^{\rm{GR}}_{i} - \frac{b_{\rm{CS}}}{2\kappa} \Pi
D_{i}\theta + \frac{a_{\rm{CS}}}{\kappa} \epsilon_{i}{}^{jk}\Bigl(
D^{l}\theta D_{j}X_{kl} \nonumber \\  &\quad - \tfrac{1}{2}
D_{j}\theta D^{l} \left( 3 E^{\rm{GR}}_{kl} + 4 X_{kl} \right)  +
A^{l}{}_{j} E^{\rm{GR}}_{kl} \Pi \nonumber \\ & \quad +
\left(E^{\rm{GR}}_{jl} + X_{jl} \right)D^{l}D_{k}\theta - \tfrac{1}{2}
D_{j} M^{\rm{GR}}_{k} \Pi \nonumber \\& \quad + \tfrac{1}{2}
A^{l}{}_{j} \left( M^{\rm{GR}}_{k} D_{l}\theta -\tfrac{1}{2}
M^{\rm{GR}}_{l} D_{k}\theta \right) \Bigr)        \nonumber\\ &\quad
+\frac{a_{\rm{CS}}}{\kappa}\Bigl( \tfrac{1}{2} D^{j} \theta \left( 3
A^{k}{}_{i} B_{jk} - A^{k}{}_{j} B_{ik} \right)  \nonumber \\ & \quad
+ B_{ij} \bigl(\tfrac{1}{3} K D^{j}\theta - D^{j}\Pi \bigr)  \Big)\,.
\end{align}
These are the model dependent constraints associated with spatial
diffeomorphism  invariance and the freedom in the foliation.
It is not clear how  standard methods for constructing
solutions to the constraints in GR could  be modified to deal with
these constraints when~$\theta$ and~$\Pi$ are  non-vanishing.  

\vspace{10pt}
\paragraph*{Evolution equations:} 
We now turn to the derivation of the time evolution equations.  Their
geometric subset provides the kinematic degrees of freedom describing
the evolution of the 3-metric $\gamma_{ij}$, the scalar field $\theta$
and the tracefree part of the extrinsic curvature~$A_{ij}$.  
They come from the definitions of the time reduction variables, 
Eqs.~\eqref{eq:ExtrCurv} and~\eqref{eq:MCSdefKtheta}, and of the Weyl
tensor, Eq.~\eqref{eq:WeylinEB}, yielding
\begin{align}
\p_{t}\gamma_{ij} &=  - 2\alpha \left( A_{ij} + \tfrac{1}{3}\gamma_{ij} K \right)
        + \Lie_{\beta}\gamma_{ij}  
\,,\nonumber\\ 
\p_{t}\theta      &= - \alpha\,\Pi + \Lie_{\beta}\theta 
\,,\nonumber\\ 
\p_{t}A_{ij}      &=  - [D_i D_j\alpha]^{\rm{TF}} 
        + \alpha\big(2 X_{ij} + E^{\rm{GR}}_{ij} 
\nonumber\\ & \quad\quad 
        - A^{k}{}_{i} A_{jk} - \frac{1}{3} \gamma_{ij} A^{kl} A_{kl} \big)
        + \Lie_{\beta} A_{ij} 
\,,
\label{eq:dtgamthetA}
\end{align}
where $E^{\rm{GR}}_{ij}$ is given in Eq.~\eqref{eq:EGR_vac}.
The previous expressions have been derived solely from
\emph{geometric} relations.
The model dependent, dynamic degrees of
freedom enter through the  EoMs yielding evolution
equations for the time reduction  variable of the scalar field, the
trace of the extrinsic curvature and  the electric part of the Weyl
tensor. They encode information about  the considered theory of
gravity including GR as well as higher derivative modifications.
Because we will employ the well-known relations in GR as abbreviations
in the following, let them serve as an example.  We recover the field
equations of GR (minimally coupled to a scalar field) if we set
$a_{\rm{CS}}=0$ and $b_{\rm{CS}}=1$ in Eqs.~\eqref{eq:dCSEoM4D}.
We find
\begin{align}
X_{ij} &=  - \frac{1}{4\kappa} [D_i\theta D_j\theta]^{\rm{TF}}\,,
\nonumber\\ \p_t K &= \alpha( \mathcal R + K^2) - D^i D_i\alpha  -
\frac{\alpha}{2\kappa} D^i\theta
D_i\theta+\Lie_{\beta}K\,,\nonumber\\ \p_t \Pi &=-D^{i}\alpha
D_i\theta-\alpha(D^{i}D_{i}\theta - K \Pi) + \Lie_{\beta} \Pi\,.
\end{align}
Together with an appropriate choice of gauge conditions 
for the lapse function and shift vector
these relations close the PDE system.
We recognize that GR essentially results in four constraints, evolution equations for~$K$ and~$\Pi$
and five algebraic relations for the electric part of the Weyl
tensor in terms of other~$3+1$  quantities. 
Due to the presence of higher derivative terms this is no longer true in dCS gravity.
Using Eqs.~\eqref{eq:dCSEoM4D}, we are still able to find evolution
equations for the time reduction variable of the scalar field and the
trace of the extrinsic curvature, which are
\begin{align}
\p_{t} \Pi & = - D^{i}\alpha D_{i}\theta - \alpha\left(
D^{i}D_{i}\theta - K \Pi \right) \nonumber\\ &\quad + 4 \alpha
\frac{a_{\rm{CS}}}{b_{\rm{CS}}} B^{ij} \left( X_{ij} +
E^{\rm{GR}}_{ij} \right) + \Lie_{\beta} \Pi \,,\nonumber\\
\label{eq:dtKdCS}
\p_{t} K & = -D^iD_i\alpha+\alpha [H^{\textrm{GR}} + A_{ij}A^{ij} +
  \tfrac{1}{3}K^2 ] +\Lie_\beta K \nonumber\\ &\quad 
- \alpha a_{\textrm{CS}} \Big[  2 A^{ij} \tilde X_{ij} + (D\times
  M^{\textrm{GR}})_iD^i\theta \nonumber\\ &\quad  - B^{ij}(A_{ij}\Pi -
  D_iD_j\theta) \Big]
- \alpha \frac{b_{\textrm{CS}}}{2\kappa} D_i\theta D^i\theta \,.
\end{align}
The relation involving the electric part of the Weyl tensor can be
derived as the trace-free contribution of the spatial projection of
Eq.~\eqref{eq:dCSEoM4D}.  This computation yields a lengthy equation
for~$\p_t\tilde{X}_{ij}$ of the form 
\begin{align}
\label{eq:EdCS}
\p_{t} \tilde{X}_{ij} \simeq & - \frac{\alpha\Pi}{L} \Big(
          D_{s} \tilde{X}_{ij}
        - 2 D_{(i} \tilde{X}_{j)s}
        - s_{(i} D^{k} \tilde{X}_{j)k}
\nonumber \\ & \quad
        -\frac{3L}{2} \epsilon_{(i}{}^{k} s_{j)} D_{k} X_{ss}
        + 3 s_{(i|} D_{s} \tilde{X}_{|j)s}
\nonumber \\ & \quad
        + \gamma_{ij} D^{k} \tilde{X}_{sk}
\Big)
+\Lie_{\beta} \tilde{X}_{ij}
\end{align}
where we present only terms corresponding to the highest spatial
derivatives of the metric.
The full equation is presented in App.~\ref{app:EdCS}.

\vspace{10pt}
\paragraph*{Closing the system:}
As we discussed in the beginning of this section, knowledge
of~$\tilde{X}_{ij}$ is not sufficient to close the PDE system, since
it  does not yield~$X_{ss}$. It is however possible to derive an
algebraic  equation for~$X_{ss}$ by projecting the tracefree part of
the spatial projection of the field equations~\eqref{eq:dCSEoM4D}
along  the gradient of the scalar field, leading to,
\begin{align}
&\left(1-\frac{3\,a_{\textrm{CS}}^2}{\kappa\,b_{\textrm{CS}}}B_{ss}^2\right)X_{ss}
  - \frac{2\,a_{\textrm{CS}}}{L\,\kappa}\Pi\,D^i\tilde{X}_{is}=  -
  \frac{b_{\textrm{CS}}}{6\,\kappa} L^2  \nonumber\\ & +
  \frac{a_{\textrm{CS}}}{3\,\kappa}\Pi\,A^{ij}B_{ij} +
  \frac{2\,a_{\textrm{CS}}}{3\,\kappa}\Pi\, K B_{ss} -
  \frac{a_{\textrm{CS}}}{\kappa}\Pi\,A^i{}_{s}B_{is} \nonumber\\ & +
  \frac{2\,a_{\textrm{CS}}^2}{\kappa\,b_{\textrm{CS}}}B^{ij}E_{ij}^{\textrm{GR}}B_{ss}
  +
  \frac{a_{\textrm{CS}}}{\kappa}(B_{s}{}^iD_{s}D_i\theta-B_{ss}D^iD_i\theta)
  \nonumber\\ & - \frac{a_{\textrm{CS}}}{3\kappa}\,B^{ij}D_iD_j\theta
  +
  \frac{2\,a_{\textrm{CS}}^2}{L\,\kappa\,b_{\textrm{CS}}}B_{ss}B_{AB}^{\textrm{TF}}\,q^{AC}\epsilon^{DB}\tilde{X}^{\textrm{TF}}_{CD}
  \nonumber\\ & -
  \frac{8\,a_{\textrm{CS}}^2}{L\,\kappa\,b_{\textrm{CS}}}B_{ss}B_{sA}\epsilon^{AB}\tilde{X}_{sB}
  +
  \frac{2\,a_{\textrm{CS}}}{\kappa}A^i{}_s\tilde{E}^{\textrm{GR}}_{is}
  \nonumber\\ & -
  \frac{2\,a_{\textrm{CS}}}{L\,\kappa}D^i\Pi\,\tilde{E}^{\textrm{GR}}_{is}
  + \frac{a_{\textrm{CS}}}{6\kappa}(D\times
  M^{\textrm{GR}})_iD^i\theta \nonumber\\ & +
  \frac{a_{\textrm{CS}}}{L\,\kappa}\Pi\tilde{A}^{i}{}_{s}M^{\textrm{GR}}_i
  + \frac{a_{\textrm{CS}}}{3\,\kappa}A^{ij}\tilde{X}_{ij} -
  \frac{2\,a_{\textrm{CS}}}{L\,\kappa}D^i\Pi\,\tilde{X}_{is}\nonumber\\ &
  + \frac{2\,a_{\textrm{CS}}}{\kappa}A^i{}_s\tilde{X}_{is} 
  - \frac{3\,a_{\textrm{CS}}}{2\,L^2\,\kappa}\Pi\,\tilde{X}^{ij}D_iD_j\theta\,.
\label{eq:Xss}
\end{align}
Besides~$X_{ss}$ this expression involves quantities for which the
time evolution  equations are known and thus this relation closes the
evolution equations, provided that we are in the generic
situation where~$3a_{\textrm{CS}}^2B_{ss}^2\ne\kappa\,b_{\textrm{CS}}$
and $D^{i}\theta\neq0$.
Once the generic case is examined we ought to treat the special
case in which this equation can not be inverted, and 
the case in which~$s^i$ is not well-defined. 

\vspace{10pt}
\paragraph*{Structure of the field equations:}
Let us begin by  assuming that we are in the generic situation, and consider
the shape  of the resulting equations. 
In short, the system does not have an~FT$N$S structure. The term breaking the
structure is the second on the left hand side of Eq.~\eqref{eq:Xss}. 
Since this term is present in the vector constraint~\eqref{eq:dCSMomentum}, one
might try eliminating it by adding multiples of the constraint. To
avoid any suspense: it is easily checked that an~FT$N$S structure is
{\it not}  recovered with this strategy. Indeed, the constraint
additionally  contains terms of the form~$D_i M^{GR}_j$,
involving higher derivatives of the metric, which do not
cancel. In order  to see explicitly the structural problem, let us
keep only terms containing the highest spatial derivative
in Eq.~\eqref{eq:Xss} and plug these into Eq.~\eqref{eq:EdCS}.  
We observe the following terms spoiling the FT$N$S structure
\begin{align}
\p_t\tilde{X}_{ij}  \sim \Pi s_{(i} \epsilon_{j)}{}^{k}D_{k}X_{ss}
\,.
\end{align}
These look diffusive and, indeed, with $X_{ss}\sim \Pi D^{i}\tilde X_{si}$, 
the highest derivative terms acting
on~$\tilde{X}_{ij}$ are given by
\begin{align}
\p_t\tilde{X}_{si}  \sim  \Pi\epsilon_{i}{}^{j}
D_{j}D_{k}\tilde{X}_s^{k}{},
\end{align}
where we have projected Eq.~\eqref{eq:EdCS}
along~$s^i$  to show exactly the problematic term. 

This second order combination does not  necessarily vanish, and 
cannot be replaced by lower derivatives using the  constraints without
introducing different higher derivative terms. Note, that the
simplifying procedure involves pushing~$s^i$ inside two derivatives,
producing  third order derivatives of~$\theta$, but these terms are
consistent with the  FT$N$S structure. The heart of the problem lies
in the fact that the operator~$\mathcal{O}$ is  not invertible. This leads to one
degree of time derivative less in one of the  equations, without
affecting the number of spatial derivatives. 
Repeated application of the operator~$\mathcal{O}$ does not allow closure of the evolution
system in a  different way because the operator~$\mathcal O$  is not nilpotent, 
as we have shown in App.~\ref{app:On}.
This means that in the generic case the system is not consistent  with the first requirement
of defining a hyperbolic system.

\vspace{10pt}
\paragraph*{Special case I:}
Let us now turn our attention to the special case for which the
coefficient  in front of $X_{ss}$ in Eq.~\eqref{eq:Xss} vanishes,
i.e.,~$3 a^2_{\rm{CS}} B^{2}_{ss} = b_{\rm{CS}}\kappa$. In this case,
it is not possible to recover~$X_{ss}$ which implies that we cannot
determine all  components of~$X_{ij}$ and the PDE system is not even
closed.

\paragraph*{Special case II:}
Now consider the case in which the gradient of the scalar
field vanishes, implying that the spatial normal vector~$s^{i}\sim
D^{i}\theta=0$  is not defined. 
Let us focus on the full field equation~\eqref{eq:EdCS_full}, 
which is the crucial relation that needs to be solved to close the system. Taking~$D_{i}\theta=0$
yields,
\begin{align}
\label{eq:EdCSDTheta0}
\p_{t}\tilde{X}_{ij} = &  \alpha\Big(
          \Pi (D\times X)_{ij}  
        - \frac{\kappa}{a_{\rm{CS}}} X_{ij}  
        - \Pi [ A^{k}{}_{(i} B_{j)k} ]^{\rm{TF}}  
\nonumber \\ & 
        + \tfrac{2}{3} K \Pi B_{ij} 
        + 2 \frac{a_{\rm{CS}}}{b_{\rm{CS}}} B_{ij} B^{kl}\left( E^{\rm{GR}}_{kl} + X_{kl} \right)
\nonumber \\ & 
        + \Pi \epsilon_{(i|}{}^{kl}  \left( \tfrac{1}{2} A_{|j)k} M^{\rm{GR}}_{l} - X_{|j)k} D_{l}\ln\alpha \right) 
\nonumber \\ & 
        + \epsilon_{(i|}{}^{kl} E^{\rm{GR}}_{|j)k} D_{l}\Pi  
\Big)
        + \Lie_{\beta} \tilde{X}_{ij}
\,.
\end{align}
Although this equation provides a prescription for the time evolution
of~$\tilde{X}_{ij}$, this tensor vanishes and we have no means to
recover~$X_{ij}$  which is required to close the PDE system. Instead
we could regard  Eq.~\eqref{eq:EdCSDTheta0} (with the left-hand side
vanishing) as differential relation for~$X_{ij}$ only. Albeit this
equation appears to be  somewhat simpler than in the generic case, we
have not found a solution to this  differential equation or a way to
use it for prescribing a time evolution  equation for $X_{ij}$. Note,
that we cannot resort to the~$2+1$ decomposition  that we employed in
the generic case, because $D_{i}\theta=0$. Thus, we have not  been
able to close the PDE system in case that the scalar field gradient
vanishes.

\paragraph*{Summary of the initial value problem:}
To setup the Cauchy problem for dCS gravity such that neither of the 
special cases above occurs initially, one requires initial data for the 
evolution variables~$\gamma_{ij},A_{ij},K,\tilde{X}_{ij}$ in the gravity sector and 
for the fields~$\theta,\Pi$ in the scalar sector
with the additional conditions that~$D_i\theta\ne0$
and~$3a_{\textrm{CS}}^2B_{ss}^2\ne\kappa\,b_{\textrm{CS}}$ everywhere.
These variables must satisfy the 
constraints~\eqref{eq:dCSHamiltonian}-\eqref{eq:dCSMomentum}. They evolve 
according to~\eqref{eq:dtgamthetA},\eqref{eq:dtKdCS} and~\eqref{eq:EdCS}. 
Note, that the variable~$\tilde{X}_{ij}$ could be replaced 
in the state vector by the electric part of the Weyl tensor~$E_{ij}$ 
according to~\eqref{eq:DefOp} using~\eqref{eq:XfromXtilde} and~\eqref{eq:Xss}.
While doing so makes the resulting expressions more cumbersome,
it does not effect the basic structure of the system.

\subsection{Well-posedness discussion}

It was previously shown in Refs.~\cite{Garfinkle:2010zx}
and~\cite{Ayzenberg:2013wua} that upon linearization around a
Schwarzschild or slowly rotating BH background,  respectively, the dCS
field equations admit superluminal mode solutions,
which are damped away. 
Studies of dCS gravity in the background of a Kerr BH revealed that the 
scalar field diverges on the inner  horizon~\cite{Konno:2014qua}. As described in
Sec.~\ref{section:Hyperbolicity}  this type of analysis is not
strong enough to draw conclusions about  well-posedness of the initial
value problem, where we are required to consider  an arbitrary
background. We also presented the structure that an FT$3$S system  of
PDEs must have in order to have a chance to be strongly hyperbolic. We
have  seen in the previous section that dCS gravity does not have this
form. This  is most evident by combining Eqs.~\eqref{eq:EdCS}
and~\eqref{eq:Xss}, in  which the operator~$\mathcal{O}$ plays the key
role. 

\vspace{10pt}
\paragraph*{Model for the structure of the dCS equations:}
For illustration consider the model equation: 
\begin{equation} O
\p_t\begin{pmatrix}u \\ v\end{pmatrix}  +
\begin{pmatrix} u \\ v\end{pmatrix} +\begin{pmatrix}  v \\ u\end{pmatrix}' = 0,
\end{equation} 
where~$O$ is the non-invertible and non-nilpotent operator, 
\begin{equation} O
= \begin{pmatrix} 1 & 0 \\ 0 & 0 \end{pmatrix} 
\end{equation} 
for definiteness. This is the situation present in dCS gravity where we should
think of~$O\,u$ as~$\tilde{X}_{ij}$ and~$O\,v$ as~$X_{ss}$. The first
equation  is a differential equation for $u$ while the second is an
algebraic relation  for~$v$. Plugging the solution for~$v$ into
the equation for~$u$ leads to 
\begin{align}
\dot u + u - u'' =& 0
\,.
\end{align}
This is the heat equation plus a non-principal term,
which does not have a FT$N$S structure.
Of course, in the dCS equations the specific form of the resulting PDE is not as simple as this,
but also does not admit a first order reduction.

\vspace{10pt}
\paragraph*{Gravitational wave degrees of freedom:}
Constructing hyperbolic formulations of systems with gauge freedom is
more  subtle than for simpler examples like the wave
equation~\cite{Hilditch:2013sba}.  Therefore one might object to the
non-existence of a first order reduction by  suggesting that the
problem is related to a poor gauge choice. One approach  might be to
try and ape the construction of the generalized harmonic  formulation
of GR, but taking~$\theta$ as the time coordinate. Since the  equation
of motion for~$\theta$ contains~$\Box\theta$, and the choice
simultaneously removes the troublesome~$D_i\theta$ terms, this
approach  initially seems promising. Unfortunately the 
``gauge source function'' would then behave as~$E_{ij}B^{ij}$, and these terms would
again take a non-hyperbolic  character. In fact, since gravitational
waves can be thought of as the  propagating part of the Weyl tensor
and the electric part of the Weyl tensor
is~$E_{ij}=E^{\textrm{GR}}_{ij}+X_{ij}$ we have shown that the lack of
hyperbolicity  in the dCS field equations,
in the generic case of $D_{i}\neq0$ and keeping the background {\textit{arbitrary}},
occurs {\it precisely} in the GW  degrees of freedom.

\vspace{10pt}
\paragraph*{Classification of dCS gravity:}
The dCS field equations admit a set of constraint and evolution PDEs.
After appropriate manipulation, their analogue in GR corresponds to a 
set of elliptic- and hyperbolic-type PDEs, respectively. 
In contrast, for dCS gravity we have seen that the PDEs encoding the propagation 
of the Weyl tensor are not hyperbolic.
How may we classify them? Since we have higher spatial derivatives the
first guess is to check for parabolicity, or perhaps a mixed
hyperbolic-parabolic structure. But since the higher derivatives do
not appear in the form~$a^{ij}\p_i\p_ju$ with~$a^{ij}$ positive
definite,  this possibility is also to be discarded, as is that of a
mixed hyperbolic-Schr\"odinger class. 
A further possibility is that the
equations may form a mixed hyperbolic-elliptic system as found 
for example in Ref.~\cite{AndMon01}. 
In such a system some variables 
appear without evolution equations, instead satisfying elliptic equations. 
The evolution equations (up to the expected freedom in the choice of lapse and
shift) in dCS gravity, however, are generically complete, and so not of this 
type. 
Since the equations seem not to lie in any 
particular PDE class, there is no definitive theory of well-posedness 
available to fall back on for the IVP. Therefore we present here a 
preliminary calculation to demonstrate what type of behavior the present 
higher derivative terms may cause, leaving a detailed study for future work. 

\vspace{10pt}
\paragraph*{Cursory mode analysis:}
Eq.~\eqref{eq:EdCS} prescribes the evolution of four degrees  of
freedom, two each in~$\tilde X_{si}$ and~$\tilde X_{AB}$. Consider the
second and first derivatives terms in these equations, by Fourier  
transforming the spatial dependence according  to~$X(t,x^i) = X(t)
\exp(i \omega\,\hat{\omega}_i x^i )$,  where~$\hat\omega_i$ is a unit
vector, that we choose to be orthogonal  to~$s^i$. Let us further
define~$\hat\nu^i$ such that~$\hat\nu$ is  orthogonal to both~$s^{i}$
and~$\hat{\omega}_{i}$. The state  vector
is~$(\tilde{X}_{s\hat\omega},\tilde{X}_{\hat\omega\hat\omega},
\tilde{X}_{s\hat\nu},\tilde{X}_{\hat\omega\hat\nu})$,
and~$X_{ss}$ is to be replaced using its equation. 
Keeping only the highest derivative terms, we get
\begin{align}
\p_t
\begin{pmatrix}
\tilde X_{s\hat\omega}\\ \tilde X_{\hat\omega\hat\omega}\\ \tilde X_{s\hat\nu}\\
\tilde X_{\hat\omega\hat\nu
} 
\end{pmatrix} 
= - i\,\omega \frac{\Pi}{L}
\begin{pmatrix}
0 &\frac{1}{2}&0&0\\  1&0&0&0\\ C_1 \omega&0&0&\frac{1}{2}\\ 0&0&1&0
\end{pmatrix}
\begin{pmatrix}
\tilde X_{s\hat\omega}\\ \tilde X_{\hat\omega\hat\omega}\\ \tilde X_{s\hat\nu}\\
\tilde X_{\hat\omega\hat\nu
}
\end{pmatrix},
\label{eq:mode_equation}
\end{align}
where 
\begin{align}
C_1 =- \frac{3\,i\,a_{\rm{CS}}\,\Pi}{L\Big(1-3\frac{a_{\rm{CS}}}{\kappa
    b_{\rm{CS}}}B_{ss}^2\Big)}\,.
\end{align}
Generically~$C_1\neq0$ since otherwise either~$a_{\rm{CS}} = 0$,
implying that the dCS modification disappears, or~$\Pi=0$. If the 
latter condition holds
everywhere this is not dynamical dCS gravity.

This equation is obtained under the simplifying assumptions that
$B_{ij},\ s^i,\ L,\ \Pi$ are constant. This approximation is justified
by the fact that PDE analysis implies freezing coefficients and
treating them as independent. 
For consistency, we should have kept~$D_i B_{jk}$ terms and~$D_i M^{\rm{GR}}_j$, 
but our aim here is just
to point out the effect that  higher derivative terms are likely to have
in this type of analysis. Except for the additional factor
of~$\omega$ inside the matrix in Eq.~\eqref{eq:mode_equation}, the 
matrix looks like the principal symbol of a weakly hyperbolic PDE. 
Computing the  general solution one finds frequency dependent growth
like~$\omega^2\,t$, so the  problem is ill-posed. 

\vspace{10pt}
\paragraph*{A final model:}
Now consider a model problem indicating the implausibility of
obtaining well-posedness of the IVP for the full dCS system. Take,
\begin{align}
\p_tu&= a\, \p_xu   + b\, \p_xv \,,\nonumber\\ \p_tv&= c\, \p^2_xu +
d\, \p_xv \,,\label{eq:final_model}
\end{align}
with~$a,b,c$ and~$d$ real constants. 
Fourier transforming in space we have an ordinary differential equation (ODE) with 
\begin{align}
\p_t\tilde{U}&= M\,\tilde{U} \,,
\end{align}
where~$\tilde{U}=(\tilde{u},\tilde{v})^T$.
Assume that~$c$ is nonzero, otherwise we are in the
standard first order case. For brevity we also assume that~$\omega>0$.
The symbol is,
\begin{align}
M&=\begin{pmatrix} a & b\\ i\,\omega\,c & d
\end{pmatrix}\,i\,\omega\,.
\end{align}
If~$b\ne0$ the Eigenvalues of the symbol are,
\begin{align}
\lambda_{\pm}&=\frac{1}{2}\,i\,\omega\,\Big[(a+d)\pm
  \sqrt{(a-d)^2-4\,i\,\omega\,b\,c}\,\Big]\,,
\end{align}
which results in mode solutions that propagate with arbitrarily fast
group  velocity and, worse, blow up in a frequency dependent
exponential manner.  Thus, the IVP is ill-posed. Assume next
that~$b=0$; then the eigenvalues of  the symbol are~$i\,\omega\,a$ 
and~$i\,\omega\,d$. If furthermore~$a\neq d$  the
symbol can be diagonalized by the similarity matrix,
\begin{align}
S=\begin{pmatrix} 1 & 0\\  \tfrac{c}{a-d}\,i\,\omega  & 1
\end{pmatrix}.
\end{align}
But as~$\omega\to\infty$, we find that~$|S|$ diverges, which prevents
application of the Kreiss matrix theorem (see Theorem~$2.4.1$
in~\cite{KreLor89})  to build estimates on solutions; the PDE is once
again ill-posed. Finally  consider the case that~$a=d$. This is
closest to what we obtained for dCS gravity in Eq.~\eqref{eq:mode_equation}.
In this case the symbol is not diagonalizable. We again  find that the system is
ill-posed, although only with growth like~$\omega^2\,t$.  Although
this seems the mildest ill-posedness, note that in the presence of
lower  order terms this growth becomes as rapid as before. In summary,
the model  problem~\eqref{eq:final_model} always has an ill-posed
IVP. Naturally one  should not use this sketch to draw conclusions
about the full dCS theory, but  in the absence of a simple model of
the same structure with a well-posed IVP,  there seems little reason
to be optimistic.

\section{Chern-Simons gravity as an effective theory}
\label{section:Effective}

It has been argued that dCS gravity can be treated as an effective field
theory~\cite{polchinsky:string,Alexander:2009tp,Adak:2008yg,Yunes:2013dva}
in which the coupling constant is treated as a parameter in a
perturbative expansion around GR.  We explore the effective-field
theory approach in two steps. 
In Sec.~\ref{ssec:reducedordermodel} we start with a ``reduced-order model'' 
suggested in Ref.~\cite{Yunes:2013dva},
in which the effective EoMs have at most second derivatives of the metric
but the dynamical variables are left arbitrary.
In Sec.~\ref{ssec:linearization} we follow the more common approach, 
see e.g. Refs.~\cite{Pani:2011gy,Yagi:2012ya,Konno:2014qua,Stein:2014xba},
and perform an order-by-order reduction, in which both the EoMs and dynamical variables
are expanded in terms of the dCS coupling parameter.

\subsection{Reduced-order model}\label{ssec:reducedordermodel}

Under the small-coupling assumption we can remove the higher
derivative terms in the EoMs~\eqref{eq:dCSEoM4D} that
prevents dCS gravity, when regarded as a ``full'' theory, from having an FT$N$S shape.
We accomplish this order-reduction by substituting the trace-reversed
form of Eq.~\eqref{eq:dCSEoM4D}  into the
C-tensor~\eqref{eq:CtensorDCS} and keeping only terms up to
$\mathcal{O}(a_{\rm{CS}})$.  With this treatment the higher derivative
terms are replaced by derivatives of the  C-tensor thus becoming a
contribution of order $\mathcal{O}(a_{\rm{CS}}^2)$ which we discard.
This procedure yields modified EoMs
\begin{align}
G_{ab} + \frac{a_{\rm{CS}}}{\kappa} C^{(2)}_{ab}  -
\frac{b_{\rm{CS}}}{2\kappa} T^{\theta}_{ab}  & = 0
\,,\nonumber\\ \Box\theta  + \frac{a_{\rm{CS}}}{4\,b_{\rm{CS}}}
\,^{\ast}\!R R & = 0 \,,
\label{eq:dCSEoMOR}
\end{align}
where the energy-momentum tensor $T^{\theta}_{ab}$ is given by
Eq.~\eqref{eq:TTtheta} and 
\begin{align}
C^{(2)}_{ab} = & \left(\nabla^{c}\nabla^{d}\theta\right)
\,^{\ast}\!W_{d(ab)c}
\end{align}
denotes the second term of the C-tensor given in
Eq.~\eqref{eq:CtensorinWeyl}.
In order to analyse the PDE structure of the order-reduced equations
of motion~\eqref{eq:dCSEoMOR} we perform a spacetime split and formulate 
them as a first order in time PDE system.  In analogy to
Sec.~\ref{section:dCS} we employ the electromagnetic decomposition  of
the Weyl tensor~\eqref{eq:EMdecompWeyl} and, in particular, we will
again use the tensor $X_{ij} = E_{ij} - E^{\rm{GR}}_{ij}$ instead of
the electric part $E_{ij}$ itself.

The kinematic evolution equations which result from geometry,
i.e. those for the 3-metric $\gamma_{ij}$,  scalar field $\theta$ and
trace-free part $A_{ij}$ of the extrinsic curvature, remain unaltered
and are given by Eqs.~\eqref{eq:dtgamthetA}.
On the other hand, the dynamic degrees of freedom given in the EoMs
determine the constraints and the evolution
equations for the momentum $\Pi$ of the scalar field,  the trace $K$
of the extrinsic curvature as well as a relation for $X_{ij}$.
Considering the order-reduced EoMs~\eqref{eq:dCSEoMOR}
we find the scalar and vector constraints
\begin{align}
\mathcal{H} = & H^{\rm{GR}} - \frac{b_{\rm{CS}}}{2\kappa} \left( \Pi^2
+ D^{i}\theta D_{i}\theta \right) \nonumber \\ & + \frac{2
  a_{\rm{CS}}}{\kappa} B^{ij} \left( \Pi A_{ij} - D_{i}D_{j}\theta
\right) \,,\nonumber \\ \mathcal{M}_{i} = & M^{\rm{GR}}_{i} -
\frac{b_{\rm{CS}}}{2\kappa} \Pi D_{i} \theta \nonumber \\ & +
\frac{a_{\rm{CS}}}{\kappa} B_{ij} \left( A^{jk} D_{k} \theta +
\tfrac{1}{3} K D^{j} \theta - D^{j} \Pi \right) \nonumber \\ & +
\frac{a_{\rm{CS}}}{\kappa} \epsilon_{i}{}^{jk}  \left( \Pi A^{l}{}_{j}
- D^{l}D_{j} \theta \right) \left( E^{\rm{GR}}_{kl} + X_{kl} \right)
\,,\label{eq:dCSconstraintOR}
\end{align}
where $H^{\rm{GR}}$ and $M^{\rm{GR}}_{i}$ are the Hamiltonian and
momentum constraints for vacuum GR given by
Eqs.~\eqref{eq:GRHamiltonian} and~\eqref{eq:GRmomentumconstraint}.
The time evolution of the scalar field momentum and the trace of the
extrinsic curvature  are prescribed by
\begin{align}
\p_{t} \Pi = & - D^{i}\theta D_{i}\alpha - \alpha \left(
D^{i}D_{i}\theta - K \Pi \right) \nonumber \\ & + 4 \alpha
\frac{a_{\rm{CS}}}{b_{\rm{CS}}} B^{ij} \left( X_{ij} +
E^{\rm{GR}}_{ij} \right) + \Lie_{\beta} \Pi \,,\nonumber \\ \p_{t} K
= & - D^{i}D_{i} \alpha + \alpha\left( H^{\rm{GR}} + A_{ij} A^{ij} +
\tfrac{K^2}{3} \right) +\Lie_{\beta} K  \nonumber \\ & + \alpha
\frac{a_{\rm{CS}}}{\kappa} B^{ij} \left( \Pi A_{ij} - D_{i}D_{j}
\theta \right) - \alpha \frac{b_{\rm{CS}}}{2\kappa} D^{i}\theta
D_{i}\theta  \,.\label{eq:dtPiTrKOR}
\end{align}
The final piece of information comes from the (trace-free part of the)
spatial projection of the EoMs. 
In contrast to the full theory the order-reduced model~\eqref{eq:dCSEoMOR} 
provides a relation {\textit{algebraic}} in $X_{ij}$, as in GR.
Keeping only terms up to $\mathcal{O}(a_{\rm{CS}})$ yields
\begin{align}
X_{ij} = & \frac{a_{\rm{CS}}}{\kappa} \Big( \tfrac{2}{3} K \Pi B_{ij}
- \Pi [ B_{k(i} A^{k}{}_{j)} ]^{\rm{TF}} \nonumber \\ &\quad + [
  B_{k(i} D_{j)}D^{k}\theta ]^{\rm{TF}}  - B_{ij} D^{k}D_{k}\theta
\nonumber \\ &\quad + \epsilon_{(i|}{}^{kl} E^{\rm{GR}}_{|j)k} \left(
D_{l} \Pi - \tfrac{1}{3} K D_{l} \theta - A_{lm} D^{m}\theta \right)
\Big) \nonumber \\ & + \frac{a_{\rm{CS}} b_{\rm{CS}}}{4\kappa^2}
\epsilon_{(i|}{}^{kl} D_{|j)} \theta D_{l} \theta \left( D_{k}\Pi -
A_{kn} D^{n}\theta \right) \nonumber \\ & -
\frac{b_{\rm{CS}}}{4\kappa} [D_{i} D_{j} \theta]^{\rm{TF}}
\,.\label{eq:XijOR}
\end{align}
We have been able to eliminate all terms involving a
coupling  between~$X_{ij}$ and the gradient of the scalar field
$\theta$. In the case  of small couplings to the dCS correction
Eq.~\eqref{eq:XijOR} closes the system  of evolution equations for any
value of the scalar field.

The relation~\eqref{eq:XijOR} involves terms that have at
most second  spatial derivatives of the metric (given by  terms~$\sim
E^{\rm{GR}}_{ij}\sim \R^{\rm{TF}}_{ij}$) and in the scalar field and
at most  first spatial derivatives of the extrinsic curvature (given
by  terms~$\sim B_{ij}=(D\times A)_{ij}$). 
This implies that the entire system of  evolution PDEs of the order-reduced dCS model given
by  Eqs.~\eqref{eq:dtgamthetA},~\eqref{eq:dtPiTrKOR} and~\eqref{eq:XijOR} 
has an~FT$N$S structure (specifically~FT$2$S),
and has a chance to be hyperbolic.
In contrast to the situation in GR, however, 
the coefficients entering the principal part do  not depend only
on the metric, but also connection terms, for example~$K$
and~$A_{ij}$.  In a full hyperbolicity analysis we must take these
coefficients to have arbitrary  values in the background solution, and
then we expect that there will be situations  in which the resulting
linearized equations are not hyperbolic. Somehow these  background
solutions will have to be disallowed by the theory, if the IVP is to
be  well-posed for all admissible initial data. 

The computation presented in this section shows that, unlike the full field 
equations, the order-reduced dCS model admits a first order reduction. 
The order-reduced field equations take the form needed for the application of the 
Cauchy-Kowalevskya theorem, as applied to Lovelock gravity in 
Ref.~\cite{ChoquetBruhat:1988dw}.
The calculation thus supports the claim that a higher-derivative gravity theory  
may be transformed into a hyperbolic system, by employing the order-reduction 
method for effective field theories (see, e.g., Ref.~\cite{Yunes:2013dva} and
references therein). However, hyperbolicity of the resulting equations will depend
crucially on the background solution under consideration. 

The key assumption underlying 
this discussion is not only a small coupling, but also  that the higher derivative 
terms in the series expansion modifying GR are at most of  the same
magnitude as the lower derivative terms, so that terms of
order~$\mathcal{O}(a^2_{\rm{CS}})$ are negligible. This is in direct
contradiction with  the approximations made in the PDEs' analysis, in
which the highest derivative terms are  taken to dominate. 
Even given initial data satisfying the condition it is not clear whether the higher  
derivative terms will remain small in the course of the evolution,
unless it is enforced explicitly by the numerical scheme.

\subsection{Small coupling expansion}
\label{ssec:linearization}
In this section we treat dCS gravity as an ``effective theory''  that
would be solved order-by-order in the perturbation parameter which is taken to be 
the coupling. Using a simple counting argument we will show that  to
\emph{every order} in the dCS coupling, i.e. to every order in the
perturbation, the EoMs can (i) be formulated as first
order in  time reduction of the theory; and (ii) have the structure of
a hyperbolic PDE system.

Let us assume that the metric and the scalar field can be expanded
according to
\begin{align}
\theta = & \sum_N c^N \theta^{(N)} \,,\quad  g_{ab} =   \sum_N c^N
g^{(N)}_{ab} \,,
\end{align}
where~$c=a_{\textrm{CS}}/\kappa$. Note, that we chose to scale the
coupling with~$\kappa$, but it can equivalently be scaled with
$b_{\rm{CS}}$.  We stress that the expansion is made over the coupling
parameter, which is formally different than a ``regular'' perturbative
approach because the small parameter of the perturbative approach
appears explicitly in the field equations. The approach we follow here
is somehow similar to that in Refs.~\cite{Konno:2014qua,Stein:2014xba} 
where the rotating BH solution in dCS gravity
is approximated with a perturbation in the coupling.  Assuming that we
know all the fields up to order $c^{N-1}$ for a given value of~$N$,
the equations for the components of the metric and scalar field at
order~$N$ are given by a linear perturbation around the background of
the metric and scalar field truncated up to order~$c^{N-1}$. 
The important point of the argument is that the basic properties of the PDE
system are encoded in the principal symbol of the equations, which
are unaffected by lower order terms in the coupling~$c$.  
These lower order terms appear as sources for the equations at
order~$c^N$.  In order to show explicitly this statement, let us first
formally expand the d'Alembertian in power of~$c$:
\begin{align}
\Box \theta \approx & \sum_{n=0}^\infty \sum_{i=0}^n
c^n\Box_{[n-i]}\theta^{(i)} ,
\label{eq:boxphi}
\end{align}
where~$\Box_{[i]} $ is the d'Alembertian truncated at order $i$ in the
expansion.  The d'Alembertian~$\Box_N$ at order $N$ is explicitly
given by
\begin{align}
\Box_N &= -\frac{1}{2} g^{(N)}{}_c{}^c\Box_0\\ 
&+\frac{1}{\sqrt{-g^{0}}}\partial_a \sqrt{-g^{(0)}}\left(g^{(N)ab}
+ \frac{1}{2}g^{(0)ab}g^{(N)}{}_c{}^c\right) \partial_b.\nonumber
\end{align}

The strategy we use is to build a linear perturbation, say $\delta
g_{ab}$ around the order $c^{N-1}$ background, say $g_{ab}$, and set
$\delta g_{ab} = c^N g^{(N)}_{ab}$, $g_{ab} = \sum_{i=0}^{N-1} c^i
g_{ab}^{(i)}$. Then, the terms contributing to $c^N$ are all of the
form $g^{(N)} g^{(0)}$. Along these lines, indices of $g_{ab}^{(N)}$
are raised and lowered with $g_{ab}^{(0)}$.

To order~$c^N$, the fields to be solved for are~$\theta^{(N)}$
and~$g^{(N)}_{ij}$.  The two terms involving these fields in
Eq.~\eqref{eq:boxphi} are~$\Box_0 \theta^{(N)}$ and~$\Box_N
\theta^{(0)}$, and their derivative structure is of the schematic form
\begin{align}
\partial^2 \theta^{(N)} + \partial g^{(N)}_{ab} & = \mbox{Source
  terms}.
\end{align}

The same reasoning applies to the gravitational equations, where the
dynamical part, coming from the Einstein tensor has the structure
\begin{align}
\partial^2 g^{(N)}_{ab} + \partial g^{(N)}_{ab} & = \mbox{Source
  terms}.
\end{align}
Finally, the terms causing the pathologies of the nonlinear theory
come from the C-tensor and are always associated with with the
coupling $c=a_{\rm{CS}}/\kappa$, reducing the overall order of such
terms by one. As a consequence, to a given order $N$, these pathological
terms are always evaluated from solutions of order $N-1$, i.e. terms
already known in an iterative scheme.

The whole argument works only if everything is well-defined to order
$c^0$, i.e. the background is a solution of GR minimally
coupled to a scalar field. Then the argument can be applied
iteratively. This is indeed the case, since to order $c^0$, the theory
is only GR with a scalar field, which is known to pose no problem.

In summary, the equations at order~$N$ have the following structure
\begin{align}
\label{eq:LinSystem0}
&\Box_0 \theta^{(N)} = \left.V'(\theta)\right|_{c=0}\theta^{(N)} +
      {\rm{l.o.t.}}\,,\nonumber\\
&\frac{1}{2}
      (\Delta^{\rm{GR}})^{(0)}_{ab}{}^{cd} g^{(N)}{}_{cd}= \frac{b_{CS}}{\kappa}
     \Bigl( \nabla^{(0)}_{(a}\theta^{(N)}
\nabla^{(0)}_{b)}\theta^{(0)}\nonumber\\
&\quad-\frac{1}{2}\nabla^{(0)}_{c}\theta^{(N)} \nabla^{(0)}{}^{c}\theta^{(0)}
g^{(0)}_{ab}\Bigr)+
            {\rm{l.o.t.}}\,,
\end{align} 
where $(\Delta^{\rm{GR}})^{(0)}$ is the operator governing
perturbations around an arbitrary background in GR, evaluated with the
order $c^0$ background,  $\nabla^{(0)}_a$ is the covariant derivative
compatible with $g^{(0)}$, and ``l.o.t.'' denotes lower order terms.

As a consequence, the principal symbol at order $c^N$ is schematically
given by
\begin{align}
\mathcal P = & \left( \begin{array}{cc} \Box_0 & 0\\ 0 &
  (\Delta^{\rm{GR}})^{(0)}
                    \end{array}\right),
\end{align}
where it is understood that $\mathcal P$ acts on $v$,
\begin{align}
v = & \left( \begin{array}{c} \theta^{(N)}\\ g^{(N)}_{ab}
           \end{array}\right).
\end{align}

In other words, in an effective approach where all the corrections in
the coupling are computed order by order, the highest order operator
decouples.  This implies that the third order derivatives always
appear only in the  source terms and dCS gravity -- when treated as an
{\textit{effective theory}} -- can be formulated as a hyperbolic set
of PDEs.

The zero-th order in the coupling trivially reduces to the Einstein
equations:
\begin{align}
G_{ab}(g^{(0)}_{cd}) &= \frac{b_{\rm{CS}}}{2\kappa}\Bigl(\nabla_a\theta^{(0)} 
\nabla_b\theta^{(0)} - \frac{1}{2}\nabla_c\theta^{(0)}  \nabla^c{}\theta^{(0)}
g^{(0)}_{ab}\nonumber\\
&\quad\quad+  V(\theta^{(0)})g^{(0)}_{ab}\Bigr),\nonumber\\ 
\Box^{(0)}\theta^{(0)} &=
  V'(\theta^{(0)}).
\end{align}

The first order in $c$ correction given by the dCS modification is
then given by
\begin{align}
&\frac{1}{2}(\Delta^{\rm{GR}})^{(0)}_{ab}{}^{cd} g^{(1)}_{cd} +
  C_{ab}(g^{(0)}_{cd}, \theta^{(0)}) 
= \frac{b_{\rm{CS}}}{2\kappa}\Bigl(2
\partial_{(a}\theta^{(1)}\partial_{b)}\theta^{(0)} \nonumber \\
&\quad -\partial_{c}\theta^{(1)}
\partial^{c}\theta^{(0)}g^{(0)}_{ab}-\frac{1}{2}\partial_{c}\theta^{(0)}
\partial^{c}\theta^{(0)}g^{(1)}_{ab}\nonumber\\
&\quad+
  V(\theta^{(0)}) g^{(1)}_{ab} + V'(\theta^{(0)}) \theta^{(1)}
  g^{(0)}_{ab}\Bigr),\nonumber\\ 
  &\Box^{(0)} \theta^{(1)} + \frac{1}{2}
  \partial_a\theta^{(0)}\nabla^{(0)}g^{(1)}_{c}{}^c =
-\frac{\kappa}{4b_{\rm{CS}}}
  \,^{\ast}\!R^{(0)}{}_{abcd}R^{(0)}{}^{abcd},
\end{align}
where  $\Box^{(0)}$ is the d'Alembertian constructed from $g^{(0)}$
only, and $C(g^{(0)},\theta^{(0)})_{ab}$ is the $C$-tensor evaluated
with the zeroth order terms of the metric and scalar field.

Higher orders become cumbersome but the structure is the same: the
term causing troubles in the non expanded theory are now evaluated on
lower order in the coupling.  In conclusion, the dynamical
Chern-Simons model, treated in this manner, can be made hyperbolic in
the same way as GR minimally coupled to a scalar field.

\section{Conclusions}\label{section:Conclusions}

We have investigated the initial value formulation and PDE structure
of dynamical Chern-Simons gravity. This modification of GR, motivated
for example by string theory, loop-quantum gravity or cosmology, 
has recently attracted a lot of attention.  
Previous studies have been concerned with the construction
of solutions   to dCS gravity and their stability properties, but
well-posedness of  the initial value problem has remained 
outstanding. We have started  filling this gap by deriving an initial
value formulation of dCS theory and investigating its PDE structure.

We encountered a number of difficulties. First, in the generic
situation when the spatial gradient of the Chern-Simons field  is
non-vanishing, if additionally~$3 a^2_{\rm{CS}}
B^{2}_{ss}=b_{\rm{CS}}\kappa$, the field equations do not close. This
means that given suitable initial  data for the
variables~$\gamma_{ij},\theta,A_{ij},K,\tilde{X}_{ij}$ and~$\Pi$   we
can not compute all components of the time derivative of the trace-free 
part of the extrinsic curvature because the electric part of the
Weyl  tensor~$E_{ij}=E_{ij}^{\rm{GR}}+X_{ij}$ is not completely
determined. Likewise when the scalar field gradient vanishes
it seems impossible to obtain the electric part  of the Weyl tensor, 
and again we can not compute the time derivative of $A_{ij}$.
To avoid either  pathology one would have to demonstrate that these cases 
can not occur.

Next, in the generic case that the spatial gradient of the scalar
field is  non-vanishing, and~$3 a^2_{\rm{CS}} B^{2}_{ss}\ne
b_{\rm{CS}}\kappa$ we succeeded  in formulating dCS gravity as an
evolution problem. But we found that the higher  derivative terms present 
in the dCS gravity EoMs have a different structure than CS
electromagnetism. A crucial tool in investigating well-posedness 
of a hyperbolic PDE system (following for 
example Refs.~\cite{Tay81,NagOrtReu04,GunGar05}) is the use of a first order reduction.
The dCS gravity EoMs do not admit such a reduction, 
and so are not hyperbolic in this sense.
Therefore one would 
naively expect that even if the dCS IVP could be made well-posed, signals could  
propagate arbitrarily fast. But, in fact, in a very rough mode analysis obtained 
by taking a subset of the full EoMs, we do not find unbounded speeds, but instead
that the IVP admits frequency dependent growth of solutions, and so is
ill-posed. The evolution PDEs of dCS gravity do not fall into any of the 
standard PDE classifications. To understand what problems the higher derivative 
terms might cause we looked systematically at a simple toy model with 
its structure inspired by dCS. The toy always  has an ill-posed initial value
problem regardless of how the various parameters  present were
chosen. It seems that further advances in PDE theory will be needed 
to make conclusive statements about the well-posedness of
the IVP of dCS gravity, but the expectation gained from the analysis of 
simplified models is that it will be ill-posed.

Perhaps anticipating this result, it has been argued that dCS gravity 
should instead be viewed as an effective model resulting from a more fundamental 
theory.  Taking on this viewpoint the dCS modifications are treated as the 
lowest order contribution in a series expansion around GR. 
We have order reduced the EoMs to eliminate the higher derivative terms yielding a 
systematically well-defined time evolution formulation. While a proper 
hyperbolicity analysis of the order-reduced PDE system is beyond the scope of this paper,
we note, that it is an FT$N$S system
and can potentially be cast into a strongly hyperbolic problem.
That said, this potential seems unlikely to be realized generically because, 
in contrast to GR, the resulting 
principal symbol contains multiple tensor fields. Somehow the field equations 
will have to disallow any ``bad'' combination of these fields. One might expect 
a similar situation in dilaton Gauss-Bonnet gravity.

We have taken the previous treatment, focusing on a series expansion
only of the EoMs, a step further and considered
perturbations of the metric and scalar field around an arbitrary
background where the expansion parameter is given by the dCS coupling
constant.  This case closely resembles most previous studies involving
dCS gravity.  We have shown that in this order-by-order expansion the
higher derivative contributions  always only enter as lower order
source terms to Einstein's equations.

Several further assumptions besides the coupling constant being a small 
underlie this computation. To justify the small-coupling assumption it has 
been argued that the dCS modification itself can be interpreted as the 
lowest order contribution to a series expansion of the underlying theory which 
would take the form $\Lie \sim \sum_{n} a^{n}\mathcal{O}(R^{n+1})$. The effective 
field theory approximation, i.e. truncating the model at~$\mathcal{O}(a)$, can 
only be valid if terms at different orders are at most comparable to each other, 
for which there is no guarantee. This assumption is particularly questionable 
in dynamical scenarios. Consider some solution to dCS gravity in the small 
coupling limit, e.g. the approximate superposition of two Schwarzschild BHs, as 
the initial configuration. One could investigate the dynamical evolution of this 
system using the Cauchy formulation of the order-reduced model, which can be 
cast into a time evolution system. 
However, near the plunge of the two BHs higher curvature modifications may 
become important,
possibly exceeding the energy cut-off,
and the small coupling approximation would break down.
This suspicion is supported by a recent study exploring highly rotating BHs  
in dCS gravity~\cite{Stein:2014xba}, 
where it has been shown that the range of validity of the perturbative approach
(considering only the dCS modification) shrinks with increasing BH spin.

Finally, thinking of the ``more fundamental'' theory as being String Theory, it
is tempting to relate the pathology in the effective theory, dCS gravity, to the
origin of the modification to GR. Recall that the dCS term derives from an 
anomaly cancellation in the gravitational sector of the 10 dimensional heterotic
string model. We argued that the anomaly cancellation procedure seems to have 
the same derivative structure as the effective dCS model. This suggests that a 
careful analysis of the anomaly canceled model should be carried out. One might 
worry about the procedure itself when the base field theory has a Lagrangian with 
the same structure as GR, though we will not enter this debate here. 

\acknowledgements

We thank V.~Cardoso for carefully proof-reading the manuscript.
We thank V.~Cardoso, T.-j.~Chen, L.~Gualtieri, E.~Lim, T.~Sotiriou, P.~Spindel, L.~Stein and N.~Yunes for many useful discussions and feedback.
H.~W. acknowledges the kind hospitality of the Yukawa Institute for Theoretical Physics at Kyoto
University during the YITP-T-14-1 workshop on
``Holographic vistas on Gravity and Strings.''
This work was supported by
the FCT--Portugal project no. CERN/FP/123593/2011, 
the FP7 ERC Starting Grant ``The dynamics of black holes: testing the limits of Einstein's theory''
grant agreement no. DyBHo--256667,
the STFC GR Rolling Grant No. ST/L000636/1, 
the NRHEP 295189 FP7-PEOPLE-2011-IRSES Grant
and the ARC contract AUWB-2010/15-UMONS-1.
D.~H. is supported in part by DFG grant SFB/Transregio~7
``Gravitational Wave Astronomy''. 
H.~W. acknowledges financial support provided under the 
{\it ERC-2011-StG 279363--HiDGR} ERC Starting Grant.

\appendix

\section{The case of $\mathcal O^n$}\label{app:On}

In this section, we show that repeated application of the
operator~$\mathcal O$  on itself never vanish, in other words that the
operator~$\mathcal O$ is  not nilpotent. If this were the case, we
could close the system by  defining a series of new fields of the
form~$X_n := \mathcal O^n \Lie_n X$ and end up with an equation
for~$\Lie_n X_n$.

We use the notation introduced in Sec.~\ref{section:dCS}. Recall
that  the operator~$\mathcal{O}$ is written as
\begin{align}
\mathcal O_{ij}{}^{kl} = L q_{(i}{}^{(l} \epsilon_{j)}{}^{k)} + L
s_{(i} s^{(l} \epsilon_{j)}{}^{k)}.
\end{align}
recall the following useful relation: 
\begin{align} 
\epsilon_{ik}\epsilon_{jl} = & q_{ij}q_{kl} - q_{ik}q_{jl}
\,.  
\end{align} 
It is a crucial remark that the term
involving a Lie derivative of the  electric part of the Weyl tensor in
Eq.~\eqref{eq:EdCS} is precisely given  by~$\mathcal{O}_{ij}{}^{kl}
E_{kl}$, giving support to the idea that  a decomposition along the
gradient of~$\theta$ is relevant.

Repeated applications of~$\mathcal O$ consists in contracting the last
two indices of $\mathcal O$ with the first two indices of the next
occurrence, e.g.  $(\mathcal O^2)_{ij}{}^{mn}=\mathcal O_{ij}{}^{kl}
\mathcal O_{kl}{}^{mn}$.

Straightforward algebra then shows that powers of~$\mathcal O$ are
given by
\begin{align}
(\mathcal
  O^{4n+2})_{ij}{}^{kl}&=\frac{1}{4^n}\,L^{4n+2}\,\left(\frac{1}{2}
  q_{ij} q^{kl} - q_{(i}^{(k} q_{j)}^{l)}\right)\,,&\quad &n\in\mathbb
  N,\nonumber\\ (\mathcal
  O^{4n})_{ij}{}^{kl}&=\frac{1}{2^{2n-1}}\,L^{4n}\,
  q_{(i}^{(k}q_{j)}^{l)}\,,&\quad &n\in\mathbb
  N^*,\nonumber\\ (\mathcal
  O^{2n+1})_{ij}{}^{kl}&=\frac{(-1)^n}{2^{n}}\,L^{2n+1}\,
  q_{(i}^{(k}\epsilon_{j)}^{l)}\,,&\quad &n\in\mathbb N^*,
\end{align}
which completes the proof.

\clearpage
\begin{widetext}
\section{Evolution equation for $\tilde{X}_{ij}$}\label{app:EdCS}
%
For completeness we present the entire time evolution equation for the 
dynamical variable $\X_{ij}$ which contains the electric part of the Weyl tensor.
We have presented its highest derivative terms in Eq.~\eqref{eq:EdCS},
highlighted in boldface in expression~\eqref{eq:EdCS_full} below.
The trace-free part of the EoMs~\eqref{eq:dCSEoM4D} fully projected onto the spatial slice
is given by
\begin{align}
-2 X_{ij} 
+ \frac{a_{\rm{CS}}}{\kappa} C^{\rm{TF}}_{ij} 
- \frac{b_{\rm{CS}}}{2\kappa} L^2 s_{i} s_{j} 
= & 0 \,,\quad{\rm{with}}\quad
C^{\rm{TF}}_{ij} = \left( \gamma^{k}{}_{i} \gamma^{l}{}_{j} - \tfrac{1}{3} \gamma_{ij} \gamma^{kl} \right) C_{kl}
\,.
\end{align}
Employing the notation 
$\Lie_{n} \tilde{X}_{ij} = \tfrac{1}{\alpha} \left(\p_{t}-\Lie_{\beta} \right) \tilde{X}_{ij}$,
the trace-free, spatial projection of the C-tensor is 
\begin{align}
C^{\rm{TF}}_{ij} = & \,
- \Lie_{n} \tilde{X}_{ij} 
+ \tfrac{3}{2} \epsilon_{(i}{}^{k} s_{j)} {\bf{\Pi\, D_{k} X_{ss}}} 
+ \Pi \left( \tfrac{2}{3} K B_{ij} - [A^{k}{}_{(i} B_{j)k}]^{\rm{TF}} \right)
- A^{k}{}_{(i} \X_{j)k} - \tfrac{1}{3} K \X_{ij}
+ A^{kl}\X_{kl} \left( s_{i}s_{j} - \tfrac{2}{3}\gamma_{ij} \right)
\nonumber \\ &\,
- 3 s_{(i} A_{j)}{}^{k} \X_{sk}
+ \X_{s(i} A_{j)s} 
- s_{(i} \X_{j)k} A^{k}_{s} 
+ 2 s_{i} s_{j} A^{s}_{k} \X_{sk}
- 3 A_{ss} s_{(i} \X_{j)s}
+ \tfrac{3}{2} \Pi \epsilon_{(i}{}^{k} s_{j)}  X_{ss} D_{k} \ln\alpha
\nonumber \\ &\,
+ \tfrac{3}{2} \Pi X_{ss} \left( 
          \epsilon_{(i|}{}^{k} D_{k} s_{|j)} 
        - \epsilon_{(i}{}^{k} s_{j)} D_{s} s_{k}
        + s_{i} s_{j} \epsilon^{kl} D_{k} s_{l} 
        \right)
+ \epsilon_{(i}{}^{k} E^{\rm{GR}}_{j)k} D_{s}\Pi
+ \epsilon^{kl} s_{(i} E^{\rm{GR}}_{j)k} D_{l} \Pi
- \epsilon_{(i}{}^{k} E^{\rm{GR}}_{j)s} D_{k} \Pi
\nonumber \\ &\,
+ \tfrac{1}{2}\Pi \left(
          \epsilon^{kl} s_{(i} A_{j)k} M^{\rm{GR}}_{l}
        - \epsilon_{(i}{}^{k} A_{j)s} M^{\rm{GR}}_{k}
        + \epsilon_{(i}{}^{k} A_{j)k} M^{\rm{GR}}_{s}
        \right)
\nonumber \\ &
+ L \Big[
-\tfrac{1}{4} s_{(i} \epsilon_{j)}{}^{k} D_{s} M^{\rm{GR}}_{k}
+ B_{k(i} D^{k} s_{j)} - B_{ij} D^{k} s_{k} 
- \tfrac{1}{3} \gamma_{ij} D^{k} B_{sk}
- \frac{1}{2} \epsilon_{(i}{}^{k} A_{j)k} \left(X_{ss} + H^{\rm{GR}} \right)
+ \frac{3}{2} s_{(i}  \epsilon_{j)}{}^{k} A_{sk} X_{ss}
\nonumber \\ &\,
+ \tfrac{1}{2} \epsilon_{(i}{}^{k} E^{\rm{GR}}_{j)l} A^{l}{}_{k}
- \tfrac{1}{2} \epsilon_{(i}{}^{k} A_{j)}{}^{l} E^{\rm{GR}}_{kl}
- \epsilon_{(i}{}^{k} E^{\rm{GR}}_{j)k} \left( A_{ss} + \tfrac{1}{3} K \right)
+ \epsilon_{(i}{}^{k} E^{\rm{GR}}_{j)s} A_{sk}
+ \tfrac{1}{2} \epsilon^{kl} A^{m}{}_{k} E^{\rm{GR}}_{lm} \left(\tfrac{1}{3} \gamma_{ij} - s_{i} s_{j} \right)
\nonumber \\ &\,
+ \tfrac{1}{2} s_{(i} \epsilon_{j)}{}^{k} \left( E^{\rm{GR}}_{kl} A^{l}_{s} - A^{l}{}_{k} E^{\rm{GR}}_{ls} \right)
- \epsilon^{kl} s_{(i} E^{\rm{GR}}_{j)k} A_{sl}
+ \tfrac{1}{4} s_{(i} \epsilon_{j)}{}^{k} D_{k} M^{\rm{GR}}_{s}
+ \tfrac{1}{4} \epsilon_{(i}{}^{k} D_{j)} M^{\rm{GR}}_{k}
+ \tfrac{1}{4} \epsilon_{(i|}{}^{k} D_{k} M^{\rm{GR}}_{|j)}
\nonumber \\ &\,
+ \tfrac{1}{2} \epsilon_{(i}{}^{k} M^{\rm{GR}}_{j)} D_{k}\ln\alpha
+ \tfrac{1}{2} \epsilon_{(i|}{}^{k} M^{\rm{GR}}_{k} D_{|j)}\ln\alpha
+ \tfrac{1}{4} \epsilon^{kl} D_{k} M^{\rm{GR}}_{l} \left( s_{i} s_{j} + \tfrac{1}{3} \gamma_{ij} \right)
\Big]
\nonumber \\ &
+ \frac{\Pi}{L}\Big[
- {\bf{ D_{s} \X_{ij} }}
+ \tfrac{1}{2} {\bf{D_{(i} \X_{j)s} }}
+ {\bf{s_{(i} D^{k} \X_{j)k}}}
- 3 {\bf{ s_{(i|} D_{s} \X_{|j)s}}}
- 2 \gamma_{ij} {\bf{ D^{k} \X_{sk} }}
+ 3 s_{i} s_{j} D^{k} \X_{sk}
\nonumber \\ &\,
+ \tfrac{1}{2} \left( 
          \epsilon_{(i}{}^{k} \X_{j)l} D_{s} \epsilon_{kl}
        + \X^{l}{}_{k} \epsilon_{(i|}{}^{k} D_{s} \epsilon_{|j)l}
        - 3 s_{(i} \epsilon_{j)}{}^{l} \X^{k}_{s} D_{s} \epsilon_{kl}
        \right)
- \tfrac{7}{2} \X_{s(i|} D_{s} s_{|j)}
\nonumber \\ &\,
- \left( \X_{ij} + 3  s_{(i} \X_{j)s} \right) D_{s} \ln\alpha 
- \tfrac{1}{2} \X_{ij} D_{k} s^{k}
+ \tfrac{1}{2} \X_{k(i} D_{j)} s^{k}
+ \tfrac{1}{2} \gamma_{ij} \X_{kl} D^{k} s^{l} 
- \tfrac{3}{2} s_{i} s_{j} \X^{k}_{s} \epsilon^{lm} D_{m} \epsilon_{kl}
\nonumber \\ &\,
+ \tfrac{1}{2} \X_{k(i} D^{k} s_{j)} 
- s_{(i} \X_{j)s} D_{k} s^{k} 
+ \tfrac{9}{2} \X_{sk} s_{(i} D^{k} s_{j)}
- 2 \epsilon_{(i|}{}^{k} \X^{l}_{s} D_{k} \epsilon_{|j)l}
\nonumber \\ &\,
+ \tfrac{1}{2} \left( s_{(i} \X^{k}{}_{j)} \epsilon^{lm} D_{l} \epsilon_{km}
        - \epsilon_{k}{}^{m} \X_{lm} s_{(i} D^{k} \epsilon_{j)}{}^{l} \right)
+ 2 \X_{s(i} D_{j)}\ln\alpha
+ s_{(i} \X_{j)k} D^{k}\ln\alpha
+ \left( 3s_{i} s_{j} -2 \gamma_{ij} \right)  \X_{sk} D^{k} \ln\alpha
\Big]
\nonumber \\ &
+ \frac{b_{\rm{CS}}\,L^3}{6 \kappa} \epsilon_{(i}{}^{k} A_{j)k}
+ \frac{a_{\rm{CS}}}{b_{\rm{CS}}} B_{ij} \Big[
 B^{kl} E^{\rm{GR}}_{kl} + 3 B_{ss} X_{ss} 
- \tfrac{2}{L} \epsilon^{kl} \left( B_{km} \X^{m}{}_{l} + 3 B_{sk} \X_{sl} \right)
\Big]
\nonumber \\ &
+ \frac{a_{\rm{CS}}\,L}{3 \kappa}\Big[
- \Pi \epsilon_{(i}{}^{k} A_{j)k} A^{lm} B_{lm}
- s_{(i} A^{k}{}_{j)} \epsilon^{lm} A_{kl} \X_{sm}
+  \epsilon_{(i}{}^{k} A_{j)l} A^{lm} \X_{km}
- 2 A_{k(i} A^{l}{}_{j)} \epsilon^{km} \X_{lm}
\nonumber \\ &\,
-  \X_{k(i} A^{l}{}_{j)} \epsilon^{km} A_{lm}
+ 3 \epsilon_{(i}{}^{k} A^{l}{}_{j)} \X_{sk} A_{sl}
-  \epsilon_{(i}{}^{k} A_{j)s} A^{l}_{s} \X_{kl}
+  \epsilon^{kl} \left( 2 \X_{sk} A_{l(i} A_{j)s} - A_{sk} \X_{l(i} A_{j)s} \right)
\nonumber \\ &\,
+ s_{(i} A_{j)k} A_{sm} \left( \epsilon^{kl} \X^{m}{}_{l} - \epsilon^{lm} \X^{k}{}_{l} \right)
- 3 A_{ss} \left( \epsilon_{(i}{}^{k} A_{j)s} \X_{sk} - s_{(i} A_{j)k} \epsilon^{kl} \X_{sl} \right)
\Big]
\nonumber \\ &
+ \frac{a_{\rm{CS}}\,L^2}{3 \kappa}\Big[
  \epsilon_{(i}{}^{k} A_{j)k} D^{l} B_{sl}
- A_{s(i} A^{k}{}_{j)} E^{\rm{GR}}_{sk}
+ s_{(i} A^{k}{}_{j)} \left( A^{l}{}_{k} E^{\rm{GR}}_{l} - A^{l}_{s} E^{\rm{GR}}_{kl} \right)
+ A_{k(i} E^{\rm{GR}}_{j)l} s^{k} A^{l}_{s}
\nonumber \\ &\,
+ A^{k}{}_{(i} A^{l}{}_{j)} E^{\rm{GR}}_{kl}
- A_{k(i} E^{\rm{GR}}_{j)l} A^{kl}
+ \tfrac{1}{2} A_{s(i|} D_{s} M^{\rm{GR}}_{|j)} 
- \tfrac{1}{2} s_{(i} A^{k}{}_{j)} D_{s} M^{\rm{GR}}_{k} 
+ \tfrac{1}{2} A^{k}{}_{(i} D_{j)} M^{\rm{GR}}_{k}
- \tfrac{1}{2} A_{k(i} D^{k} M^{\rm{GR}}_{j)}
\nonumber \\ &\,
+ \tfrac{1}{2} s_{(i} A_{j)k} D^{k} M^{\rm{GR}}_{s}
- \tfrac{1}{2} A_{s(i} D_{j)} M^{\rm{GR}}_{s}
\Big]
\,.\label{eq:EdCS_full}
\end{align}
\end{widetext}

\bibliographystyle{h-physrev4}
\bibliography{dcs}
\end{document}